\documentclass[prd, twocolumn, nofootinbib, tightenlines,preprintnumbers,notitlepage,longbibliography,superscriptaddress]{revtex4-1}

\usepackage{graphicx}
\usepackage{amsmath}
\usepackage{amsfonts}
\usepackage[colorlinks=true,citecolor=blue,urlcolor=cyan,linkcolor=black]{hyperref}
\usepackage{tablefootnote}
\usepackage{multirow}
\usepackage{relsize}
\usepackage{xcolor}
\usepackage[normalem]{ulem}
\usepackage{aas_macros}

\newcommand{\avery}[1]{\textcolor{red}{(Avery: #1)}}
\newcommand{\peter}[1]{\textcolor{blue}{(Peter: #1)}}

\newcommand{\fnl}{f_{NL}}
\newcommand{\fnlloc}{f_{NL}^{\mathrm{loc}}}
\newcommand{\fnldelta}{f_{NL}^{(\Delta)}}
\newcommand{\deltafid}{\Delta^{\mathrm{fid}}}

\newcommand{\bng}{b_{\mathrm{NG}}}

\newcommand{\unitrm}[1]{\,\mathrm{#1} }

\newcommand{\uiuc}{Illinois Center for Advanced Studies of the Universe \& Department of Physics, University of Illinois at Urbana-Champaign, Urbana, IL 61801, USA}

\begin{document}

\title{Probing beyond local-type non-Gaussianity with kSZ tomography}

\author{Peter~Adshead}
\email{adshead@illinois.edu}
\affiliation{\uiuc}

\author{Avery~J.~Tishue}
\email{atishue@illinois.edu}
\affiliation{\uiuc}

\begin{abstract}
Non-Gaussianity of the primordial curvature perturbations may arise from a variety of well motivated early Universe scenarios. In particular, inflationary theories with additional light degrees of freedom can generate a bispectrum that is peaked in the squeezed limit. While the presence of an additional massless scalar  can produce local-type primordial non-Gaussianity, in general the squeezed limit of the bispectrum depends on the mass of the new degree of freedom, and can deviate from the local shape. The resulting bispectrum leaves a distinct imprint on the amplitude and scale-dependence of the galaxy bias, which for massive fields can differ from the $k^{-2}$ scaling from local type non-Gaussianity, providing an observational window into the physics of the early Universe.  
In this work we demonstrate that kinematic Sunyaev-Zeldovich tomography with next generation cosmological surveys will offer significant additional constraining power for both the shape and amplitude of scale-dependent bias arising from primordial non-Gaussianity beyond the local type. We show that this improved constraining power is robust to various obstacles such as the optical depth degeneracy, photometric redshift errors, and uncertainty in the galaxy bias model. With CMB S4 and the Large Synoptic Survey Telescope, we forecast that compared to the galaxy survey alone the addition of kSZ tomography will offer a roughly factor of two reduction in the measurement uncertainty of the amplitude $\fnl$ of primordial non-Gaussianity well beyond the local (massless) limit. We find that kSZ tomography extends the range of masses for which order unity constraints on $\fnl$ are achievable, as well as extending the range of masses for which the late time probes of the matter power spectrum outperform the sensitivity of the CMB itself.  
\end{abstract}

\maketitle

\section{Introduction \label{sec:intro}}
Probing non-Gaussianity of the primordial density fluctuations is a key objective for upcoming next-generation cosmic microwave background (CMB) experiments, such as the Simons Observatory \cite{2018arXiv180807445T} and CMB-S4 \cite{Abazajian:2019eic, CMB-S4:2016ple}, and large scale structure (LSS) surveys, such as SPHEREx \cite{SPHEREx:2014bgr}. Detailed measurements of the distribution of structures in the Universe  have the potential to provide critical information about the mechanism responsible for setting the initial state of the Universe. On the one hand, existing observations are consistent with an early period of accelerated expansion---inflation.  A Gaussian, red-tilted and adiabatic spectrum of initial density fluctuations, consistent with models for single-field, slow-roll inflation \cite{Planck:2018jri}, accounts for all observations to date.  On the other hand, primordial non-Gaussianity holds the potential to provide insights beyond the single-field slow-roll inflationary paradigm, and probe the microphysics of the very early Universe \cite{Bartolo:2004if, Komatsu:2009kd, Alvarez:2014vva}.

One primary goal for upcoming searches is to improve measurement sensitivity to the amplitude of local-type non-Gaussianity \cite{Komatsu:2001rj}. Here the non-Gaussian primordial curvature perturbation $\zeta(\mathbf{x})$ is generated by the fluctuations in a Gaussian field $\zeta_g(\mathbf{x})$ via
\begin{align}
\zeta(\mathbf{x}) = \zeta_g(\mathbf{x}) +\fnlloc(\zeta_g(\mathbf{x})^2 - \langle \zeta_g(\mathbf{x})^2\rangle).
\end{align}
The corresponding momentum space bispectrum 
\begin{align}
  \langle \zeta(\mathbf{k}_1)\zeta(\mathbf{k}_2) \zeta(\mathbf{k}_3) \rangle' =& \fnlloc P_{\zeta}(k_1) P_{\zeta}(k_2) + \text{perms.},
 \end{align}
is peaked in the squeezed limit, $k_2,k_3 \gg k_1$. The prime, $\langle \ldots \rangle'$, denotes that we have suppressed the momentum conserving delta function $(2\pi)^3 \delta(\mathbf{k}_1 + \mathbf{k}_2 + \mathbf{k}_3 )$, which restricts the momenta to form a closed triangle. Local-type non-Gaussianity leads to mode coupling between short- and long-wavelength fluctuations. The short wavelength modes control the local density of collapsed objects like galaxies, which are then non-trivially correlated across long distances due to this type of non-Gaussianity.

The squeezed limit of the bispectrum is particularly important because single-field slow-roll inflation, or more generally single-clock inflation, predicts that the observable bispectrum should vanish in this limit \cite{Pajer:2013ana}. Consistency relations for large-scale structure \cite{Creminelli:2013poa, Creminelli:2013nua} then ensure that properties of correlation functions of galaxy overdensities are determined by local physics \cite{Creminelli:2013nua}. These relations prohibit the development of the long range correlations predicted by local-type non-Gaussianity, leaving any such correlation as a clean signature of primordial non-Gaussianity. An important benchmark, then, is the limit $\sigma (\fnlloc) \lesssim 1$  \cite{SPHEREx:2014bgr}. Reaching this limit would test the single-clock inflationary paradigm; detection of $\fnlloc \neq 0$ would rule out all single clock  inflationary models and indicate the presence of additional light degrees of freedom during inflation \cite{Creminelli:2004yq}.

While the importance of testing local primordial non-Gaussianity is manifest, in general  the details of the primordial non-Gaussianity can be complicated and need not be described by a squeezed bispectrum with amplitude $\fnlloc$. Additional light degrees of freedom during inflation \cite{Chen:2009we, Chen:2009zp}, particle production during inflation \cite{Anber:2012du}, models with higher derivative interactions \cite{Chen:2006nt,Silverstein:2003hf}, excited initial states \cite{Holman:2007na}, curvaton models \cite{Linde:1996gt}, inhomogeneous reheating \cite{Dvali:2003ar, Kofman:2003nx}, transient violations of slow-roll  \cite{Chen:2006xjb, Chen:2008wn, Flauger:2010ja, Adshead:2011bw, Adshead:2011jq, Miranda:2012rm, Adshead:2012xz, Adshead:2013zfa} and bounce cosmologies \cite{Koyama:2007if} are a few key examples of early Universe physics that can produce different types of primordial non-Gaussianity, for example predicting signals dominated by equilateral ($k_1 \approx k_2 \approx k_3$) and folded ($k_1 \approx 2k_2 \approx 2k_3$) bispectra configurations. Tests of the details of primordial non-Gaussianity are therefore tests of the details of early Universe physics, and thus have to potential to shed light on the fundamental interactions that seeded the initial conditions of the hot Big Bang. Although Planck \cite{Planck:2019kim} already constrains local and equilateral primordial non-Gaussianity to $\fnlloc = -0.9 \pm 5.1 $ and $f_{NL}^{\mathrm{equil.}} = -26 \pm 47 $ respectively, the theoretical target of order unity constraints remains to be reached. Novel and improved probes of primordial non-Gaussianity therefore remain an acutely important avenue to testing fundamental physics. 

To date, information from the CMB has provided the most stringent constraints on non-Gaussianity \cite{Planck:2013wtn, Planck:2015zfm, Planck:2019kim, Sohn:2024xzd}. However, due to the lack of remaining modes to be measured in the  primary CMB, future information will come from other probes, such as large scale structure \cite{Alvarez:2014vva}, including line intensity mapping \cite{Camera:2013kpa,MoradinezhadDizgah:2018zrs,MoradinezhadDizgah:2018lac,Liu:2020izx}, and the secondary CMB \cite{Hill:2013baa,Schmittfull:2017ffw,Munchmeyer:2018eey}. Extracting information about primordial correlation functions from these measurements remains an ongoing challenge. One promising approach is to capitalize on the wealth of information stored in the correlations induced between the CMB anisotropies and LSS observables as photons from the surface of last scattering interact with matter in the late Universe. Cosmological inference with these correlations will become increasingly powerful in the near future as next generation CMB and LSS experiments probe small scale fluctuations with unprecedented sensitivity. This includes measurements of weak lensing (see e.g. Ref. \cite{Lewis:2006fu}), the moving-lens effect \cite{1983Natur.302..315B,1986Natur.324..349G,Hotinli:2020ntd,Hotinli:2021hih}, and the thermal and kinematic Sunyaev-Zeldovich (kSZ) effects \cite{Zeldovich:1969ff,1970A&A.....5...84Z,1980ARA&A..18..537S,1972CoASP...4..173S,Sazonov:1999zp}.

In this work, we focus on the kSZ effect---a secondary CMB temperature anisotropy due to inverse scattering of photons off electrons with bulk line-of-sight velocity. The imprint of the kSZ effect on the CMB, combined with a LSS survey on the same portion of the sky, enables the reconstruction of the radial velocity field, $v_r$. In particular, the cross correlation of the radial velocity field with the galaxy field is projected to be measured with a high-signal-to-noise ratio on large scales, and can be described by an estimator that is unbiased by other secondary CMB anisotropies \cite{Smith:2018bpn}. The velocity field contains information about the density fluctuation and the growth rate. Reconstructing $v_r$ with the kSZ effect therefore provides an additional measurement of the large scale matter power spectrum that, crucially, has a different bias than reconstruction of the density field with galaxy surveys. This also facilitates sample variance cancellation, where the comparison of the density field reconstruction with two different tracers — here, galaxies and velocities —  enables improved measurement of the respective biases and the underlying matter power spectrum on large scales, which otherwise would hinder either probe in isolation due to survey volume limitations. Altogether, this makes kSZ tomography a powerful probe of cosmology and any new physics imprinted on the matter power spectrum, or the respective biases, on large scales.  As we focus on here, one such imprint is scale-dependent galaxy bias arising from primordial non-Gaussianity \cite{Dalal:2007cu, LoVerde:2007ri, Matarrese:2008nc, Shandera:2010ei, Desjacques:2011mq, Adshead:2012hs, DAloisio:2012ioh}. The scale-dependence and amplitude of this bias depends on the details of early Universe physics.  Importantly, for a variety of models, the scale-dependent bias has the most significant effect on the largest scales, which is precisely the regime where kSZ tomography is projected to have the greatest signal to  noise.

The potential of sample variance cancellation \cite{Seljak:2008xr}, as well as the additional constraining power on large scales motivates the study of kSZ tomography as a probe of primordial non-Gaussianity. Indeed, in Ref. \cite{Munchmeyer:2018eey} it was shown that kSZ tomography can be a powerful probe of local non-Gaussianity in the form of constraints on the bispectrum amplitude $\fnlloc$, improving on constraints from galaxies alone by a factor of a few. Similarly, kSZ tomography may also improve constraints on local primordial non-Gaussianity in terms of the trispectrum amplitude $\tau_{NL}$ \cite{AnilKumar:2022flx}. Furthermore, current and upcoming galaxy surveys can constrain primordial non-Gaussianty beyond the local type via measurement of scale-dependent bias on linear and quasi-linear scales \cite{Norena:2012yi,Gleyzes:2016tdh, Green:2023uyz}. 

In this paper, we carry out forecasts to show that kSZ tomography can be leveraged similarly to improve constraints on primordial non-Gaussianity beyond the local type due to its imprint as a scale-dependent galaxy bias on large cosmological scales.  The improved constraining power from kSZ tomography is due to it providing an additional measurement of the matter power spectrum, which facilitates sample variance cancellation and helps mitigate various degeneracies between the non-Gaussianity parameters and astrophysical and cosmological parameters. We also demonstrate that the additional constraining power provided by kSZ in this context is robust when including various realistic effects including photometric redshift (photo-$z$) errors and uncertainty in details of the galaxy bias, galaxy-electron correlation, and $\Lambda$CDM parameters. {We find kSZ tomography will have significant practical application for constraining primordial non-Gaussianity beyond the local type, both in terms of outperforming the CMB and for achieving the theoretical target of order unity constraints. } In our analysis we show this practical application applies not only to pushing down upper limits on the amplitude of primordial non-Gaussianity, but also to improving the measurement significance of non-zero $f_{NL}$, should it be detected in future data. 


This paper is organized as follows. In Sec.~\ref{sec:nongaussianity} we provide a brief overview of the primordial non-Gaussianity scenario that we consider and the corresponding signal in the galaxy bias. In Sec.~\ref{sec:tomography} we briefly review  galaxy and kSZ tomography, and in Sec.~\ref{sec:forecasts} we describe the setup for our forecasts. Results for our baseline forecasts are presented in Sec.~\ref{subsec:baseline_forecasts}, and the sensitivity of these results to various model and survey details is presented in Sec.~\ref{subsec:robustness}. We conclude with a discussion of these results and directions for future work in Sec.~\ref{sec:discussion}.

\section{Non-Gaussianity and Galaxy Bias \label{sec:nongaussianity}}
For simplicity, and to demonstrate the efficacy of kSZ tomography, we focus on a scenario in which the primordial non-Gaussianity arises from additional light degrees of freedom during inflation, for example in models such as quasi-single-field inflation \cite{Chen:2009we, Chen:2009zp}. When extra, light degrees of freedom are present,  the three-point function of the curvature perturbation $\zeta$ can receive contributions in the squeezed limit, which take the form
\begin{align}
    \lim_{\mathbf{k}_1 \to 0}  \langle \zeta(\mathbf{k}_1)\zeta(\mathbf{k}_2) \zeta(\mathbf{k}_3) \rangle' =& \fnldelta \left( \frac{k_1}{k_2} \right)^{\Delta} P_{\zeta}(k_1) P_{\zeta}(k_2) 
     \label{eq:bispec}
\end{align}
where the amplitude is controlled by the parameter $\fnldelta$, and the shape is controlled by the scaling exponent $\Delta_{\pm} = 3/2 \pm \sqrt{9/4 - m^2/H^2}$, where $m$ is the mass of the light field. This scenario has been explored in significant detail, see, for example, Refs. \cite{Chen:2009we,Chen:2009zp,Baumann:2011nk, Arkani-Hamed:2015bza,Green:2023uyz}.  The squeezed limit is dominated by $\Delta_-$, so that a massless field leads to local primordial non-Gaussianity, $\Delta = 0$. Further, $\Delta$ smoothly approaches $3/2$ as $m \rightarrow 3H/2$. Larger values of $\Delta$ are possible, for example a very massive spectator can yield a correction to the bispectrum with $\Delta=2$ \cite{Green:2023uyz}. Further, in single-field inflationary scenarios the leading physical term in squeezed bispectrum has a factor $\sim \! (k_1 / k_{2})^2$  \cite{Maldacena:2002vr,Creminelli:2004yq,Creminelli:2013cga} corresponding to $\Delta = 2$ (with an amplitude of the order of $\mathcal{O}(\epsilon_H))$ \cite{Green:2023uyz}. 
In what follows, we  focus our analysis on the range $\Delta \in [0,2]$.

The bispectrum in Eq.~(\ref{eq:bispec}) yields a correlation between the variance $\sigma_R$ of curvature perturbations on some scale $k$ and the primordial curvature perturbation on some scale $k^\prime$, which in turn induces a correlation between the variance and the late time total matter overdensity $\delta$,
\begin{equation}
    \langle \sigma_R(\mathbf{k}) \delta(\mathbf{k}^\prime) \rangle' \propto \frac{\fnldelta (kR)^{\Delta}}{k^2T(k)} P_{mm}(k), 
\end{equation}
where $T(k)$ is the transfer function (normalized such that $T(k) \rightarrow 1$ as $k\rightarrow 0$). Hence the galaxy overdensity on a scale $k$ is correlated with the matter overdensity on a scale $k^\prime$. The upshot is that this results in an additional, scale-dependent term in the galaxy bias, $\bng$, that takes the form \cite{Schmidt:2010gw,Desjacques:2011jb,Desjacques:2011mq,Giannantonio:2011ya,Baumann:2012bc}\footnote{Note the factor of 3 gives a normalization  $3 f_{NL}^{(\Delta = 0)} = f_{NL}^{\mathrm{loc}} $. }
\begin{equation}
    \bng (k,z) = 3 \fnldelta \frac{b_{\phi}(z)}{k^2 \mathcal{T}(k,z)} (k R_{\star})^\Delta.
    \label{eq:bngdefn}
\end{equation}
Here, $\mathcal{T}(k,z) = 2T(k)D(z)/3H_0^2 \Omega_m$, where the growth function $D(z)$ is normalized such that $D(z) \sim (1+z)^{-1}$ during matter domination. The parameter $R_{\star}$ is the comoving Lagrangian radius of the halos under consideration, determined by the minimum halo mass. We take the minimum halo mass to be $10^{13} M_{\odot}$ corresponding to $R_{\star} = 2.66 h^{-1} \unitrm{Mpc}$ \cite{Gleyzes:2016tdh}. The non-Gaussian bias parameter $b_{\phi}(z)$ can be non-trivial in general, and in principle introduces a perfect degeneracy with the non-Gaussianity amplitude $f_{NL}^{(\Delta)}$ that we hope to measure. Following Ref.\ \cite{Desjacques:2016bnm}, we assume $b_{\phi}$ is given by universality conditions so $b_{\phi}(z) = 2\delta_c (b_{1}(z) - 1)$ where $\delta_c = 1.42$ is the critical overdensity of collapse and $b_{1}(z)$ is the standard linear galaxy bias. We note that uncertainty in $b_{\phi}$, and in the universality conditions that relate $b_{\phi}$ to the linear bias $b_1$, may degrade constraints on $\fnldelta$ \cite{Barreira:2020ekm,Barreira:2021ueb,Barreira:2022sey}, although it has been shown that multitracer analyses may help circumvent this obstacle \cite{Barreira:2023rxn}. 

\section{Galaxy and kSZ tomography} \label{sec:tomography}

In linear theory, which is valid on large scales, the continuity equation relates the velocity mode to the density mode  via 
\begin{align}
v(\mathbf{k}) = i \frac{faH}{k} \delta_m(\mathbf{k}), \label{eq:continuity}
\end{align}
where 
\begin{align}
D(k,a) = \left(\frac{P_{mm}(k,a)}{P_{mm}(k,a=1)}\right)^{1/2}
\end{align}
and
\begin{align}
f(k,a) = \frac{d \ln(D(k,a))}{d \ln(a)}
\end{align}
are the growth function and the growth rate, respectively. Reconstruction of the velocity field therefore provides a measurement of the underlying density field. In particular, if the signal to noise of this reconstruction is sufficiently high, the factor of $k^{-1}$ in Eq.~(\ref{eq:continuity}) indicates that velocity tomography can potentially probe the matter power spectrum on the largest scales more precisely than a `direct' density tracer such as galaxies. 

One way to reconstruct these large-scale velocity flows is with the kSZ effect, whereby CMB photons scatter with electrons present in the gas in galaxies and galaxy clusters. Electrons with bulk peculiar-velocity $\mathbf{v}$ upscatter CMB photons via inverse Compton scattering, inducing correlations between the CMB anisotropies and the intervening large scale structure.  In the so-called `snapshot geometry' \cite{Smith:2018bpn}, in which the Universe is taken  to be a 3D box of comoving side length $L$ taken at a `snapshot' time $t$ corresponding to a redshift $z$ and comoving distance $\chi(z)$, a minimum variance quadratic estimator for the reconstructed large-scale (radial) velocity field, $\hat{v}_r$, can be written as an integral over small scales containing the observed power spectrum of the density tracer, e.g. galaxies, the observed CMB temperature power spectrum, and the galaxy-electron cross spectrum $P_{ge}$ \cite{Smith:2018bpn}. The reconstruction noise power spectrum for the radial velocity estimator is given by
\begin{align}
   \frac{1}{ N_{\hat{v}_r \hat{v}_r}(k_L,\mu, z)} = &\frac{K^2(z)}{\chi^2(z)} \times \nonumber \\
   &\int \frac{dk_Sk_S}{2\pi}  \frac{P_{ge}^2(k_S,z)}{P_{gg}^{\mathrm{obs}} (k_S,k_L,\mu,z) C^{TT,\mathrm{obs}}_{\ell = \chi(z) k_S}}, \label{eq:kSZnoise}
\end{align}
where $\mu = \hat{k} \cdot \hat{r}$, $k_L$ is the amplitude of the three-dimensional Fourier wavevector corresponding to the reconstructed large scale mode,
and the integral over small scales $k_S$ is dominated roughly by the range $k_S \in [0.1, 10] \unitrm{Mpc}^{-1}$ for the surveys we consider in this work. Here 
\begin{align}
K(z) = \sigma_T n_{e,0} x_{e}(z) e^{-\tau(z)} (1+z)^2,
\end{align}
is the kSZ radial weight function, where $\sigma_T$ is the Thompson scattering cross section, $n_{e,0}$ is the electron number density, $x_{e}(z)$ is the free electron fraction, and $\tau(z)$ is the optical depth to redshift $z$. The velocity mode and its radial component are related via $\hat{v} = \mu^{-1} \hat{v}_r$, and hence $N_{\hat{v} \hat{v}} = \mu^{-2} N_{\hat{v}_r \hat{v}_r}$. 

The observed galaxy power spectrum is 
\begin{align}
P_{gg}^{\mathrm{obs}}(k_S,k_L,\mu,z) = P_{gg}(k_S,z) + N_{gg}(k_L,\mu,z), 
\end{align}
where the noise,
\begin{align}
    N_{gg}(k_L,\mu,z) = \frac{1}{n_g(z) W^2(k_L,\mu,z)},
\end{align}
is determined by the shot-noise via the galaxy number density $n_g(z)$ (in $\mathrm{Mpc}^{-1}$) and by photometric redshift errors described by the  photo-$z$ kernel function,
\begin{align}
    W(k_L,\mu,z) = \exp{ \left(-\frac{\mu ^2 k_L^2 \sigma_z^2}{2H^2(z)} \right)}.
\end{align} 

With the galaxy field and the reconstructed radial velocity field, we have access to the galaxy and velocity power spectra, $P_{gg}$ and $P_{\hat{v}_r \hat{v}_r}$, and the cross-spectra $P_{g \hat{v}_r}$, defined in the usual way via $\langle X_{\mathbf{k}_1} Y_{\mathbf{k}_2} \rangle = (2\pi)^3 \delta( \mathbf{k}_2 + \mathbf{k}_2) P_{XY}$,
\begin{align}
    P_{gg}(k,z) &= b_g^2(z,k)  P_{mm}(k,z) \label{eq:Pgg}, \\ 
    P_{g \hat{v}_{r}}(k,z) &= b_v(z) \mu \frac{f(k)aH}{k}b_g(k,z) P_{mm}(k,z) \label{eq:Pgv} , \\ 
    P_{\hat{v}_{r} \hat{v}_{r}}(k,z) &= b_v^2(z) \mu^2 \left(\frac{f(k)aH}{k}\right)^2 P_{mm}(k,z) \label{eq:Pvv}. 
\end{align}
Here, $b_g$ is the galaxy bias and $b_v$ is the velocity reconstruction bias. The velocity reconstruction bias arises if the fiducial model for the small scale galaxy-electron cross correlation $P_{ge}(k_S)$ does not match the true underlying one, in which case the estimator for the reconstructed velocity field is biased with respect to the true field, $\hat{v}_{r} = b_v v_{r}$, where the bias $b_v$ is given by \cite{Smith:2018bpn}
\begin{align}
b_{v} &= \frac{\int dk_S F(k_S) P^{\mathrm{true}}_{ge}(k_S)}{\int dk_S  F(k_S) P^{\mathrm{fid}}_{ge}(k_S)}\,, 
\end{align}
where 
\begin{align}
F(k_S) &= k_S \frac{P^{\mathrm{fid}}_{g e}(k_S)}{P^{\mathrm{obs}}_{g g}(k_S) C_{\ell = k_S \chi}^{TT, \mathrm{obs}} }\, .
\end{align}
This is important because the small scale galaxy-electron cross correlation $P_{ge}(k_S)$ is determined largely by astrophysics and is a source of theoretical uncertainty. A precise prediction of $P_{ge}(k_S)$ requires precise knowledge of the mean optical depth of the measured galaxies \cite{Battaglia:2016xbi,Flender:2016cjy,Soergel:2017ahb}, as well as the spatial distribution of electrons \cite{Smith:2018bpn}. This presents an obstacle to using kSZ tomography for cosmological inference: for example, when attempting to use kSZ tomography to measure the large-scale galaxy-velocity cross correlation, the underlying signal is a squeezed bispectrum of two galaxy modes and one CMB mode \cite{Smith:2018bpn}, obeying $\langle \delta_g \delta_g T \rangle \propto P_{ge}(k_S) P_{gv}(k_L)$. Hence the large scale galaxy-velocity cross-spectrum we hope to measure suffers a perfect overall amplitude degeneracy with the unknown small-scale galaxy-electron cross-spectrum. This is the so called `optical depth degeneracy.' In forecasts, the optical depth degeneracy can be fully accounted for by marginalizing over $b_v$. The situation is analogous to marginalizing over the galaxy bias when attempting to use the galaxy power spectrum to constrain cosmology. Importantly, the bias $b_v$ is scale-independent, in contrast to the scale-dependent bias $\bng$ induced by primordial non-Gaussianity which we aim to constrain, and hence the optical depth degeneracy in principle does not introduce a fatal degeneracy for our purposes.

The total galaxy bias is 
\begin{align}
    b_g(k,z) =& \,\, b_1(z) + \bng(k,z) + b_{\mathrm{rsd}}(z) f \mu^2 \nonumber \\
    &+ b_{k^2}(z) (kR_{\star})^2 + b_{k^4}(z) (kR_{\star})^4,
    \label{eq:bg_ofk_full}
\end{align}
where we have included the contributions from the standard linear bias $b_1(z)$; the contribution from non-Gaussianty, $\bng(k,z)$, given by Eq.~(\ref{eq:bngdefn}); the Kaiser effect  RSD term \cite{Kaiser:1987qv} $f \mu^2$ with a bias $b_{\mathrm{rsd}}$ that accounts for anisotropic assembly bias from selection effects \cite{Obuljen:2020ypy}\footnote{In subsamples of the Baryon Oscillation Spectroscopic Survey Data Release 12 \cite{Obuljen:2020ypy,BOSS:2012dmf,BOSS:2011sdu,BOSS:2015zan}, $|b_{\mathrm{rsd}}-1|$ was shown to be roughly $0.1$ to $0.3$.}; and  gradient biases $b_{k^2}$ and $b_{k^4}$ that account for non-linear clustering on small scales. 

This model for the galaxy bias as well as the model for the observed matter power spectrum could be further extended, for example by considering a general bias expansion in the effective field theory of large scale structure (from which the gradient terms originate). However, such extensions should not change the main conclusions of this work because kSZ tomography adds constraining power on the largest cosmological scales, but additional terms arising from the bias expansion become important on small scales compared to the regime where kSZ adds appreciable cosmological information. Including the gradient terms allows us to still broadly capture the general impact that small scale contributions to the total galaxy power spectrum have on our results.

\begin{figure}
    \centering
\includegraphics[width=1.\columnwidth]{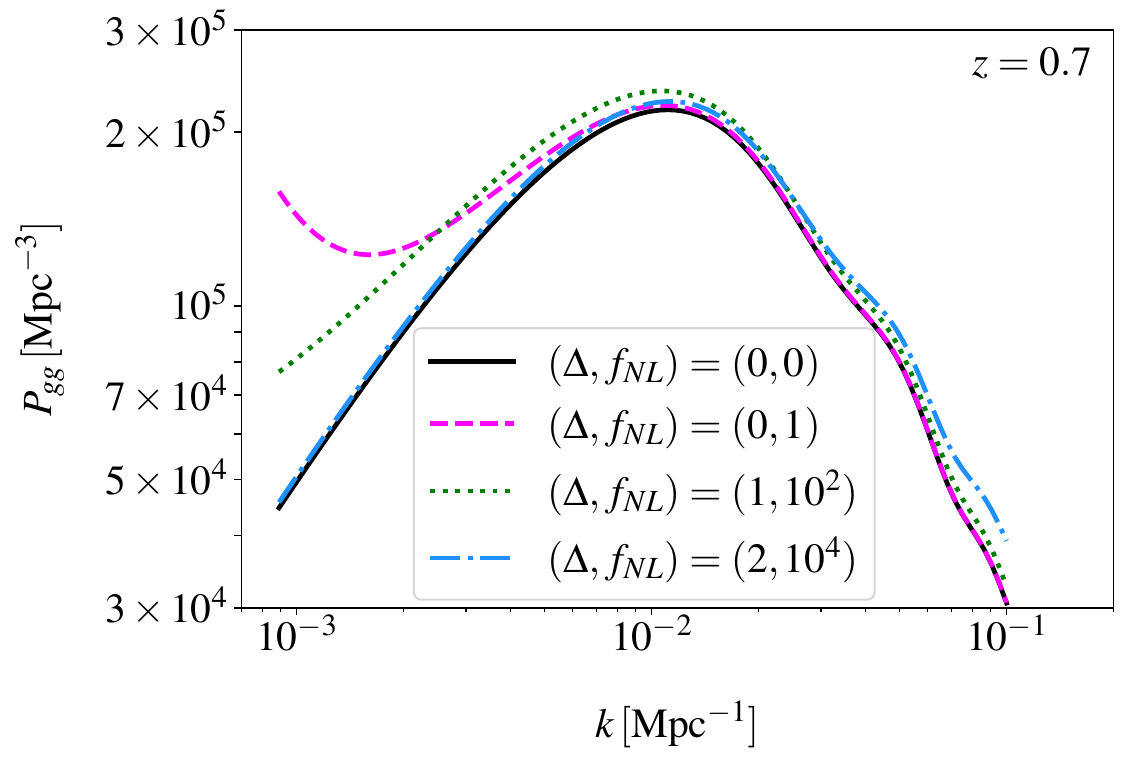}
    \caption{\textbf{Effect of scale-dependent bias from primordial non-Gaussianity on the linear galaxy power spectrum at $z=0.7$.} We show the scale-dependent modification to the linear galaxy power spectrum due to $\bng(k)$ for various pairs of scaling exponents $\Delta$ and non-zero amplitudes $\fnldelta$, compared to the case without any primordial non-Gaussianity (black, solid). All curves assume $b_1 =1.37$, $b_{\mathrm{rsd}} = 1$, and $b_{k^{2n}} = 0$. For illustrative purposes, we have selected $\fnldelta$ values to make the modification to $P_{gg}$ more visible.}
    \label{fig:Pgg}
\end{figure}

In Fig.~\ref{fig:Pgg}, we show the effect that $\bng(k)$ from primordial non-Gaussianity has on the galaxy power spectrum for a range of scaling exponents and amplitudes. For smaller $\Delta$ values, the galaxy power spectrum is enhanced on large scales where the kSZ tomography signal-to-noise is greatest. As $\Delta \rightarrow 2$, the effect on large scales diminishes and the enhancement shifts to small scales. Therefore, we naively expect kSZ to assist more in reducing $\sigma(\fnldelta)$ for smaller scaling exponents.

\section{Forecasts \label{sec:forecasts}}
We forecast results for a galaxy survey with specifications matching those proposed for the Large Synoptic Survey Telescope (LSST) Y10 \cite{LSSTScience:2009jmu} and an S4-like CMB survey \cite{CMB-S4:2016ple}. We consider five redshift bins $z \in [0.2, 0.7, 1.3, 1.9, 2.6]$, with corresponding fiducial values for the  linear galaxy bias $b_1(z) \in [1.05,1.37,1.79, 2.22, 2.74 ]$ (see Secs. 3.7, 3.8, and 13.5 in Ref.~\cite{LSSTScience:2009jmu}). For all redshift bins we take the fiducial values $b_{\mathrm{rsd}}(z) =1$, $b_{k^2}(z) = b_{k^4}(z) = 0 $, and $b_v(z) = 1$. 

To carry out our forecasts, we construct a Fisher information matrix, 
\begin{align}
    F_{ab} = \frac{V}{2} \int^{1}_{-1} d\mu\, \int \frac{k^2 dk}{4\pi^2} \, \mathrm{Tr} \left[ C^{-1}\frac{dS}{d\pi_a} C^{-1}\frac{dS}{d\pi_b} \right].
    \label{eq:fisher}
\end{align}
Here $V$ is the survey volume at each redshift bin,
\begin{align}
    V = [5.2, 43.6, 75.9, 89.3, 119.9] \times 10^{9} \, \mathrm{Mpc}^3,
\end{align}
$C_{ij}=S_{ij}+N_{ij}$ is the covariance matrix of the signals $S_{ij}$ and noises $N_{ij}$ where the indices label the observables (i.e. $S_{ij}$ corresponds to galaxy and velocity power and cross spectra $P_{ij}$ in Eqs.~(\ref{eq:Pgg})-(\ref{eq:Pvv})), $\pi_c$ is a vector of the model parameters, and the trace runs over the observable indices $ij$. For notational brevity we have suppressed the $(k,\mu,z)$ dependence and $i,j$ indices of $C$ and $dS$ in Eq.~(\ref{eq:fisher}). 
The diagonal noise matrix is 
\begin{align}
    N_{ij}(k,\mu,z) = \mathrm{diag}[N_{gg}(z),N_{\hat{v}_{r} \hat{v}_{r}}(k,\mu,z)], 
\end{align}
and the galaxy shot noise $n_{g}(z)$ in our redshift bins is taken to be \cite{Munchmeyer:2018eey}
\begin{align}
    n_{g}=\left[5,2,0.6,0.15,0.03 \right] \times 10^{-2}\, \mathrm{Mpc}^{-3}.
\end{align}
For the model of the total observed CMB $C_{\ell}^{TT,\mathrm{obs}}$ used to compute the velocity reconstruction noise, we follow Ref. \cite{Hotinli:2021hih}, including the moving lens effect, late-time and reionization kSZ contributions, the cosmic infrared background, the thermal SZ effect, radio sources, and weak lensing. We also include a white noise contribution
\begin{equation}
    N_{\ell}^{\mathrm{white}} = \Delta_T^2 \, \mathrm{exp} \left[ \frac{\ell (\ell+1) \theta_{\mathrm{FWHM}}^2 }{8 \ln(2)} \right],
\end{equation}
where $\Delta_T$ is the white noise RMS and $\theta_{\mathrm{FWHM}}$ is the beam full-width half-maximum. We assume a CMB experiment with five frequency bins, $[39,93,145,225,280] \, \mathrm{GHz}$, with corresponding beam widths $[5.1,2.2,1.4,1.0,0.9]'$ and noise RMS values $[12.4,2,2,6.9,16.6] \, \mu\mathrm{K}'$. We assume $\ell_{max}=6500$, which is a reasonable estimate for the maximum CMB multipole for which foreground modeling is expected to be under control. Later, in Sec.~\ref{subsubsec:surveys}, we further comment on how our results depend on the details of the CMB experiment.

In our forecasts, the parameters we consider are the bias prefactors appearing in Eq.~(\ref{eq:bg_ofk_full}), where in $\bng$ this corresponds to marginalizing over $\fnldelta$ and, for some of our analysis, $\Delta$. In our baseline forecasts, we also marginalize over a six parameter  $\Lambda$CDM model,
\begin{align}
    \pi^{\Lambda\mathrm{CDM}} = [A_s, n_s, H_0 [\,\mathrm{km\,s}^{-1}\mathrm{Mpc}^{-1}], \Omega_b h^2, \Omega_c h^2, \tau]
\end{align}
with corresponding fiducial values  
\begin{align}
    \pi^{\Lambda\mathrm{CDM}}_{\mathrm{fid}}=  [2.15 \times 10^{-9}, 0.9625, 0.67 , 0.022,0.12,0.066 ] \label{eq:LCDMfid}
\end{align}
Later in Sec.~\ref{subsubsec:effect_of_cosmology}, we  examine the impact that uncertainty in the $\Lambda$CDM parameters has on the forecasted constraints on primordial non-Gaussianity. 

To compute the Fisher information matrix in Eq.~(\ref{eq:fisher}), the integral over wavenumber  is an integral over the large scale modes of the galaxy and velocity spectra. This integral runs from $k_{\mathrm{min}} = \pi / V^{1/3}$, determined by the box size of the snapshot geometry, to a $k_{\mathrm{max}}$ value jointly determined by (a) the smallest scale assumed accessible by the galaxy survey, and (b) the minimum of the $k_S$ values over which we integrate to reconstruct the large scale velocity modes. In practice, we take this to be $k_{\mathrm{max}} =0.1\, \mathrm{Mpc}^{-1}$ for our baseline forecasts.  

Following the approach taken in Ref. \cite{Green:2023uyz}, we consider two cases for the non-Gaussianity parameters $\fnldelta$ and $\Delta$ in our forecasts. Current data are consistent with vanishing $\fnl$, so in the first case  we marginalize over $\fnldelta$, assume a fiducial value $\fnldelta = 0$, and take the scaling exponent to be fixed to its fiducial value, $\deltafid$, in the range $[0,2]$. This strategy is reasonable as long as the data comfortably permit $\fnldelta=0$. The reason we do not marginalize over $\Delta$ in this case is that there is no information about the shape of $b_{NG}(k,z)$ if its fiducial amplitude is assumed to be vanishing. This is immediately clear if one considers how the signals $P_{ij}$ change as $\Delta$ is varied, for which $dP_{ij}/d\Delta = 0$ for the fiducial choice $\fnldelta  = 0$. Hence, instead we ask how much can kSZ improve constraints around $\fnldelta = 0$ as the scaling exponent $\Delta$ is fixed at different values.

We also consider a second case in which we assume a non-vanishing fidicual value for $\fnldelta$ and also marginalize over the scaling exponent. This case allows us to consider how well galaxy and kSZ tomography can simultaneously measure the shape and amplitude of the scale-dependent galaxy bias if future data were to detect non-zero $\fnldelta$. This approach may not prove necessary in data analysis in the immediate future, but should hints of primordial non-Gaussianity be revealed in next-generation surveys, the analysis here elucidates some of the possibilities and challenges for simultaneous inference of the shape and amplitude of scale-dependent bias from primordial non-Gaussianity.

\section{Results \label{sec:results}}

\subsection{Baseline forecasts \label{subsec:baseline_forecasts}}
\begin{figure}
    \centering
\includegraphics[width=1.\columnwidth]{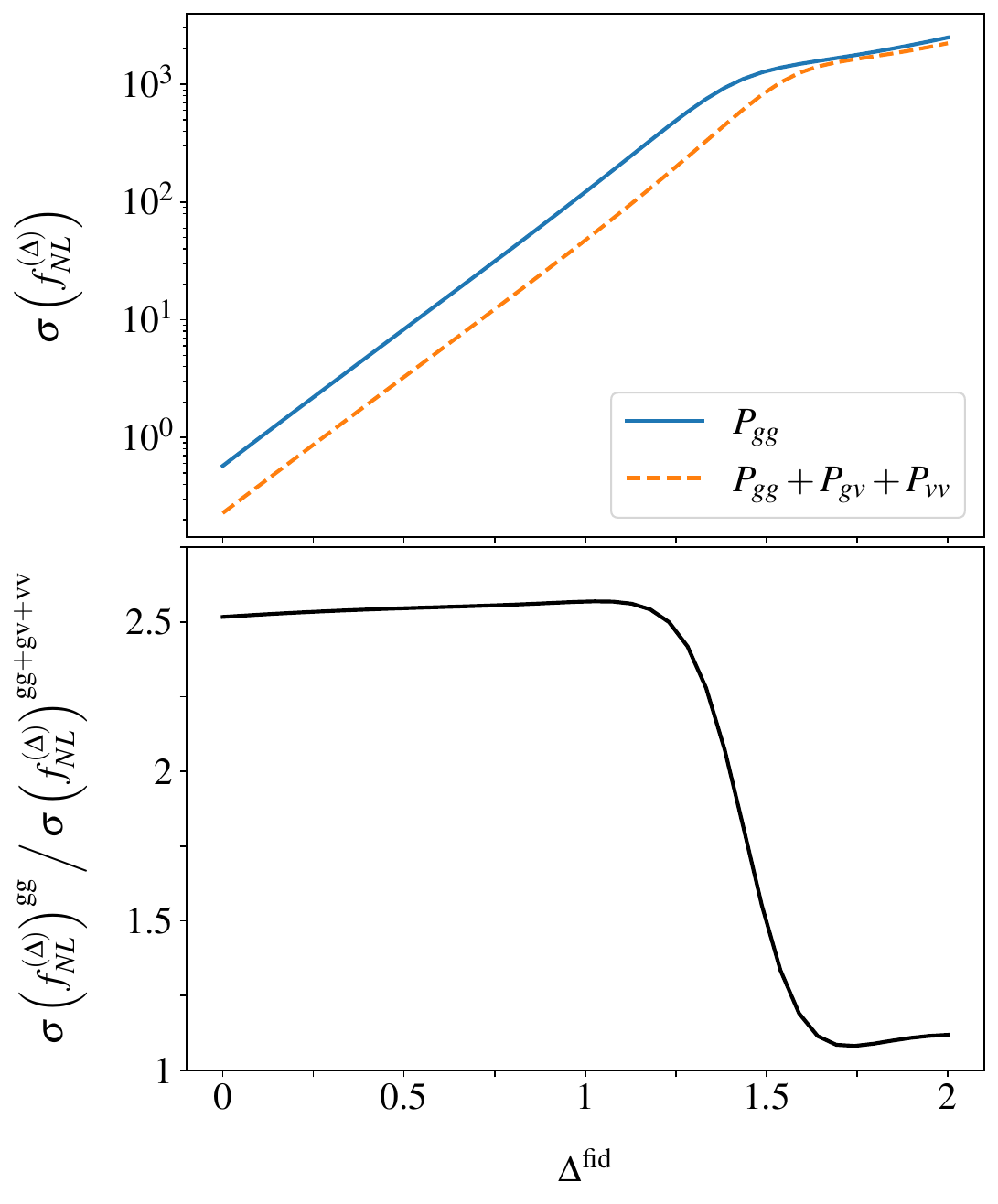}
    \caption{\textbf{Improved constraints on $\fnldelta$ due to kSZ tomography for fixed $\Delta$.} 
    We show the forecasted $1\sigma$ uncertainty on $\fnldelta$ with galaxy tomography alone (blue, solid) compared to galaxy and kSZ tomography (orange, dashed) over a range of fixed $\Delta$ values in the upper panel, and the ratio of the two in the lower panel. For each $\deltafid$ value, we take a fiducial value of $\fnldelta = 0$.}
    \label{fig:ksz_vs_gal_fnl_fixedDelta}
\end{figure}

We begin by presenting results for our baseline forecasts described in the previous section. In Fig.~\ref{fig:ksz_vs_gal_fnl_fixedDelta} we demonstrate how $1\sigma$ constraints on $\fnldelta$ improve with the inclusion of kSZ tomography in the case that $f_{NL}^{(\Delta), \mathrm{ fid}} =0$ and the scaling exponent is fixed to a value $\Delta \in [0,2]$. We first note that we find excellent agreement between our results and other work for local-type non-Gaussianity, $\deltafid=0$ \cite{Munchmeyer:2018eey}.\footnote{To compare with Ref.~\cite{Munchmeyer:2018eey}, we repeated the analysis only marginalizing over $\fnldelta$, $b_1$, and $b_v$, taking the other bias parameters and the $\Lambda$CDM parameters fixed to their fiducial values. After accounting for the factor of 3 normalization in Eq.~\ref{eq:bngdefn}, we find $\sigma(\fnlloc)^{\mathrm{gg}} = 1.39$, agreeing with Ref.~\cite{Munchmeyer:2018eey} to within $3\%$. We find $\sigma(\fnlloc)^{\mathrm{gg+gv+vv}} = 0.68$, and Ref.~\cite{Munchmeyer:2018eey} finds $\sigma(\fnlloc)^{\mathrm{gg+gv+vv}}= 0.45$. This difference is a consequence of our more conservative model for the CMB spectrum which yields a larger velocity reconstruction noise. When implementing a similar CMB spectrum as in Ref.~\cite{Munchmeyer:2018eey}, we recover their value of $\sigma(\fnlloc) = 0.45$.}  For local-type non-Guassianity, we find the inclusion of kSZ tomography yields $\sigma(\fnlloc) = 0.68$, more than two times smaller than that achievable with galaxies alone, which would permit a roughly $1.5\sigma$ exclusion of the theoretical single-field target of $\fnlloc \lesssim \mathcal{O}(1)$. 

Importantly, we see that kSZ tomography also has the ability to significantly improve constraints on the amplitude of primordial non-Gaussianity beyond $\Delta=0$, compared to galaxies in isolation. In the range $\deltafid \lesssim 1.3$, the improvement is over a factor of 2 and is weakly dependent on $\deltafid$. Importantly, the projected measurement uncertainty  $\sigma(\fnldelta)$ remains less than order one for $\deltafid \lesssim 0.3$, suggesting promising prospects for constraining primordial non-Gaussianity beyond the local type. For galaxies alone, the range for order one constraints on $\fnldelta$ is smaller, roughly $\deltafid \lesssim 0.1$. For larger scaling exponents, roughly $\deltafid \gtrsim 1.3 $, the improvement in $\sigma(\fnldelta)$ from kSZ tomography rapidly decreases, already dropping below $ \approx \! 50\%$ by $\deltafid =1.5$. When the scaling exponent is assumed to be perfectly known, kSZ tomography offers very little additional constraining power to galaxy tomography as $\Delta \rightarrow 2$. The basic reason for this is that kSZ tomography has poor signal to noise as $k$ approaches smaller (quasilinear) scales, which is where larger scaling exponents cause greater modification to the galaxy bias. Near $\deltafid =2$ the $1\sigma$ constraints on $\fnldelta$ are $\mathcal{O}(10^3)$, which is over an order of magnitude larger than current constraints on $f_{NL}^{\mathrm{equil.}}$ from Planck 2018 \cite{Planck:2019kim}. We can also compare with searches for quasi-single-field shapes in Planck data \cite{Planck:2013wtn, Sohn:2024xzd}, the most recent of which give much weaker constraints than galaxy plus kSZ tomography at $\Delta=0$,   $\sigma(f_{NL}^{\Delta=0}) =12$, but a much stronger constraint for $\Delta \gtrsim 0.85$ (or $\deltafid \gtrsim 0.65$ when compared to galaxies alone), for example yielding $\sigma(f_{NL}^{\Delta=3/2}) =26$. By way of further comparison, forecasted constraints from the CMB alone, with CMB-S4 and low-$\ell$ Planck, predict $\sigma(\fnlloc) = 1.8$ and $\sigma(f_{NL}^{\mathrm{equil.}}) = 21.2$ \cite{CMB-S4:2016ple}. These constraints are about a factor of two improvement over Planck alone, and so to first approximation we may expect the quasi-single field constraints to improve similarly, in which case galaxy and kSZ tomography would outperform the CMB alone for $\Delta \gtrsim 0.7$ (or $\approx 0.5$ without kSZ). In short, this suggests galaxy and kSZ tomography will play an important role in constraining non-Gaussianity beyond the local type, achieving the target of order unity constraints for quasi-single field shapes with scaling exponents close to the local type, roughly for $\deltafid \lesssim 0.3$, and extending the range of scaling exponents for which the probes of the late time power spectrum outperform the CMB alone. Conversely, neither galaxy nor velocity tomography will likely play a significant role in constraining larger scaling exponents in next generation surveys. 

We now turn to the second case, in which we consider how well the  amplitude and scaling exponent of the bispectrum can be inferred simultaneously if a non-zero $\fnldelta$ were detected. We follow a similar approach as in Ref. \cite{Green:2023uyz}, setting the fiducial value of $\fnldelta$ to the value consistent with a $5\sigma$ detection, $f_{NL}^{(\Delta), \rm{fid}} = 5\sigma(\fnldelta)$. However, for simplicity, and to maintain consistency with present bounds from the CMB, we impose this relation using the $\sigma(\fnldelta)$ values from Ref. \cite{Sohn:2024xzd}. We find that using the fiducial $\fnldelta$ values from this procedure yields forecasted measurement errors from galaxy and kSZ tomography that would give a $< 5\sigma$ detection in those observables internally when marginalizing over $\Delta$. In other words, quasi-single-field amplitudes $\fnldelta$ presently consistent with the CMB to within $5\sigma$ are not large enough to separately yield a $\geq 5 \sigma$ detection with galaxy and kSZ tomography if the scaling exponent is simultaneously marginalized. A minor exception occurs at very small scaling exponents, $\deltafid \lesssim 0.1 $, for which the forecasted measurement errors with galaxy and kSZ tomography are small enough to yield a $\lesssim 5.9 \sigma $ detection.

\begin{figure}
    \centering
\includegraphics[width=1.\columnwidth]{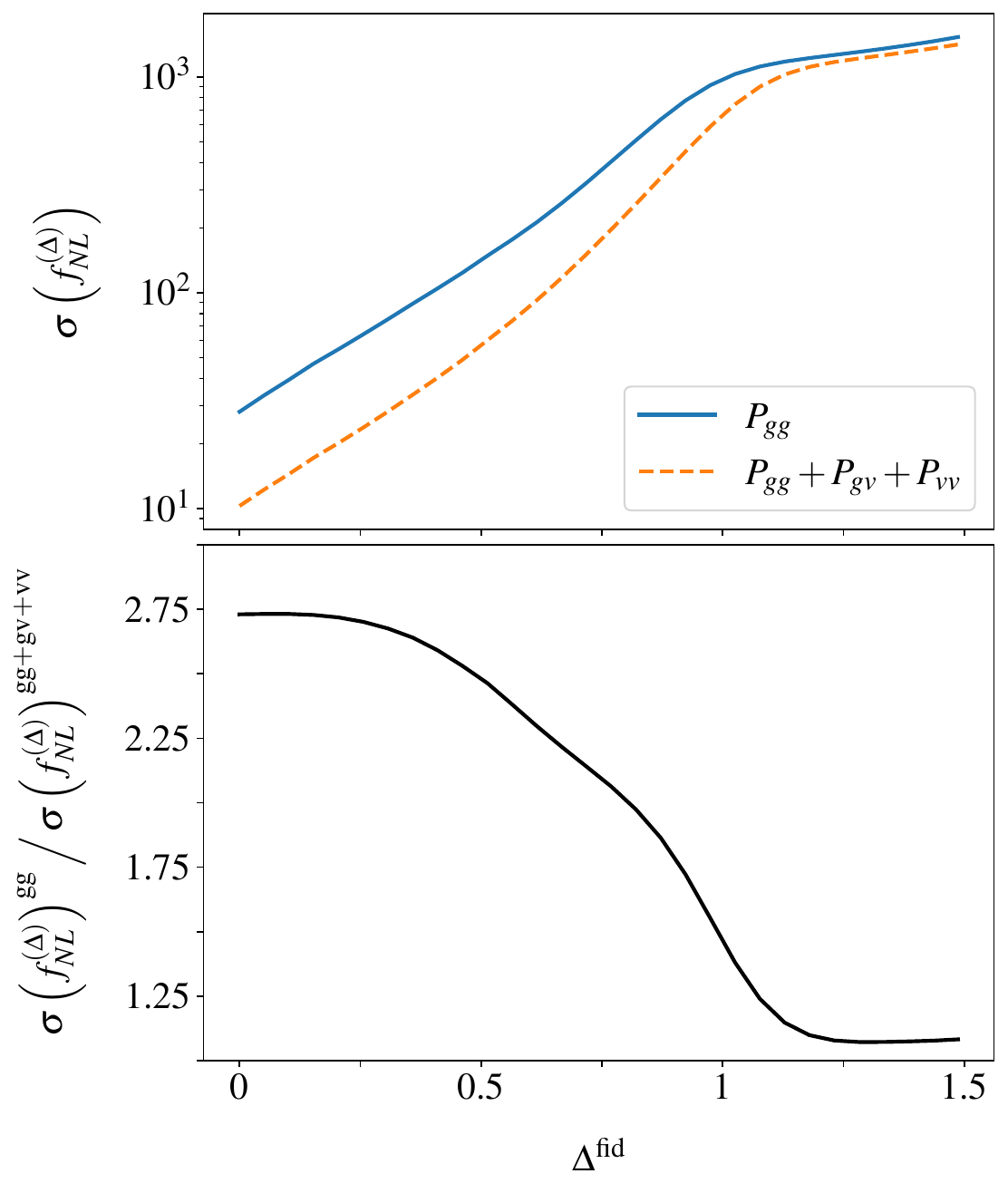}
    \caption{\textbf{Improved constraints on $\fnldelta$ with kSZ tomography for marginalized  $\Delta$.} 
    We show the forecasted $1\sigma$ uncertainty on $\fnldelta$ with galaxy tomography alone (blue, solid) compared to galaxy and kSZ tomography (orange, dashed) over a range of fiducial $\Delta$ values in the upper panel, and the ratio of the two in the lower panel. For each $\deltafid$, the fiducial amplitude for $\fnldelta$ is set as described in the main text.}
    \label{fig:ksz_vs_gal_fnl_varyDelta}
\end{figure}

\begin{figure}
    \centering
\includegraphics[width=1.\columnwidth]{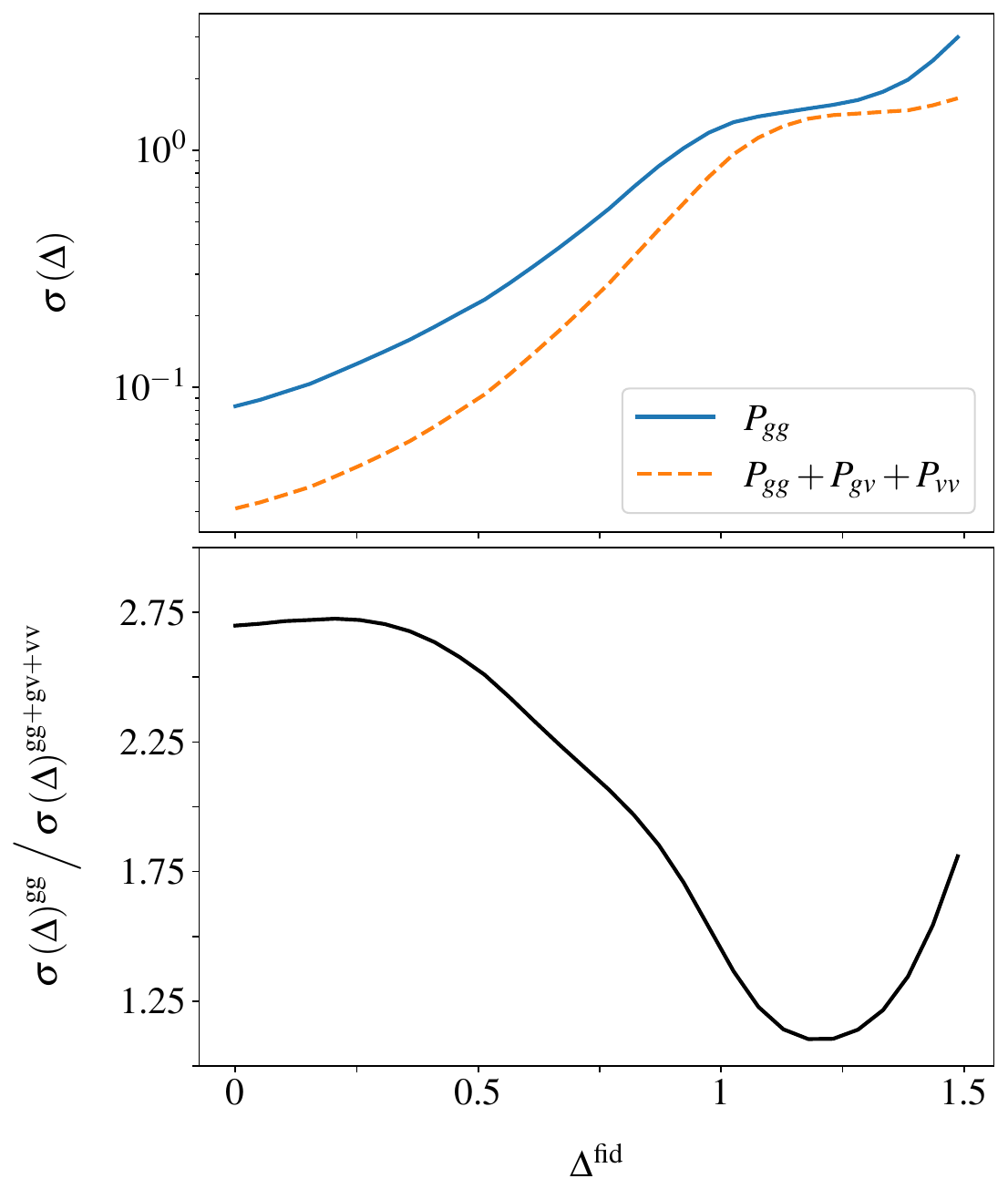}
    \caption{\textbf{Constraints on $\Delta$ with galaxy and kSZ tomography.} 
    Similar to Fig.~\ref{fig:ksz_vs_gal_fnl_varyDelta}, but instead showing constraints on the scaling exponent $\Delta$.}
    \label{fig:ksz_vs_gal_Delta_varyDelta}
\end{figure}

We show our results for this analysis in Figs.~\ref{fig:ksz_vs_gal_fnl_varyDelta} and Figs.~\ref{fig:ksz_vs_gal_Delta_varyDelta}, which show the forecast constraints on the amplitude and scaling exponent, respectively. Focusing on Fig.~\ref{fig:ksz_vs_gal_fnl_varyDelta}, we find that the $1\sigma$ measurement error on $\fnldelta$ will be significantly larger in this case than what we found in Fig.~\ref{fig:ksz_vs_gal_fnl_fixedDelta} where the scaling exponent was fixed. This is most pronounced for local-type non-Gaussianity, $\Delta=0$, where the measurement error is roughly a factor of $50$ larger, while at the effect becomes less pronounced as $\Delta$ approaches larger values, however, the constraints there are too weak to yield a significant detection with galaxy and kSZ tomography anyway. Regardless, we find that the addition of kSZ tomography can significantly reduce the measurement error on $\fnldelta$ from galaxies alone --- by over a factor of 2 --- for a range of scaling exponents. As a concrete example, for $\deltafid = 0.2$, galaxies alone would have sensitivity for a roughly $1.5\sigma$ hint, while with the addition of kSZ tomography, this would jump to a $4\sigma$ hint. Similarly,  for $\deltafid=0$, galaxies alone would yield just over a $2\sigma$ hint, while the addition of kSZ tomography would raise this to nearly $6\sigma$ detection. This highlights that kSZ tomography can play a practical role in reinforcing measurements from the CMB of primordial non-Gaussianity beyond the local type, even where the constraining power from galaxy surveys alone is limited. 

Another promising picture emerges in Fig.~\ref{fig:ksz_vs_gal_Delta_varyDelta}, where we show the forecasted measurement error on the scaling exponent from galaxy and kSZ tomography. Under the assumptions we have made here, we find that kSZ tomography will significantly improve the ability to infer the shape of the scale-dependent bias originating from light fields. Once again, this is especially relevant for smaller scaling exponents, where kSZ reduces the measurement error by over a factor of 2; this will make $\mathcal{O}(10^{-1})$ constraints on the scaling exponent possible up to $\deltafid \approx 0.5$, while such constraining power with galaxies alone is essentially only possible for local type non-Gaussianity.


\subsection{Sensitivity to model and survey details \label{subsec:robustness}}

In this section we examine how robust the kSZ-driven improvement in constraints on primordial non-Gaussianity is to various biases, degeneracies, and survey details. We focus our analysis on the case where the scaling exponent is fixed and constraints are forecasted around $f_{NL}^{(\Delta),\mathrm{fid}} = 0$. We omit an extended discussion of $b_{\mathrm{rsd}}$, but we note that we found fixing $b_{\mathrm{rsd}}$ had insignificant effect on the results; doing so improved $\sigma(\fnldelta)$ by at most $\mathcal{O}(10\%)$ for galaxy tomography alone, and only $1-3\%$ when kSZ tomography is included. The minimal impact of $b_{\mathrm{rsd}}$ is expected, given that it only enters the total galaxy bias with very mild scale-dependence via $f(k)$, which is very distinct from the scaling in $\bng(k)$. Therefore, while kSZ tomography does help mitigate the degeneracy between $b_{\mathrm{rsd}}$ and $\fnldelta$, in practice uncertainty in $b_{\mathrm{rsd}}$  has very little impact on constraints on primordial non-Gaussianity.



\subsubsection{Optical depth degeneracy \label{subsubsec:bvbias}}
\begin{figure}
    \centering
\includegraphics[width=1.\columnwidth]{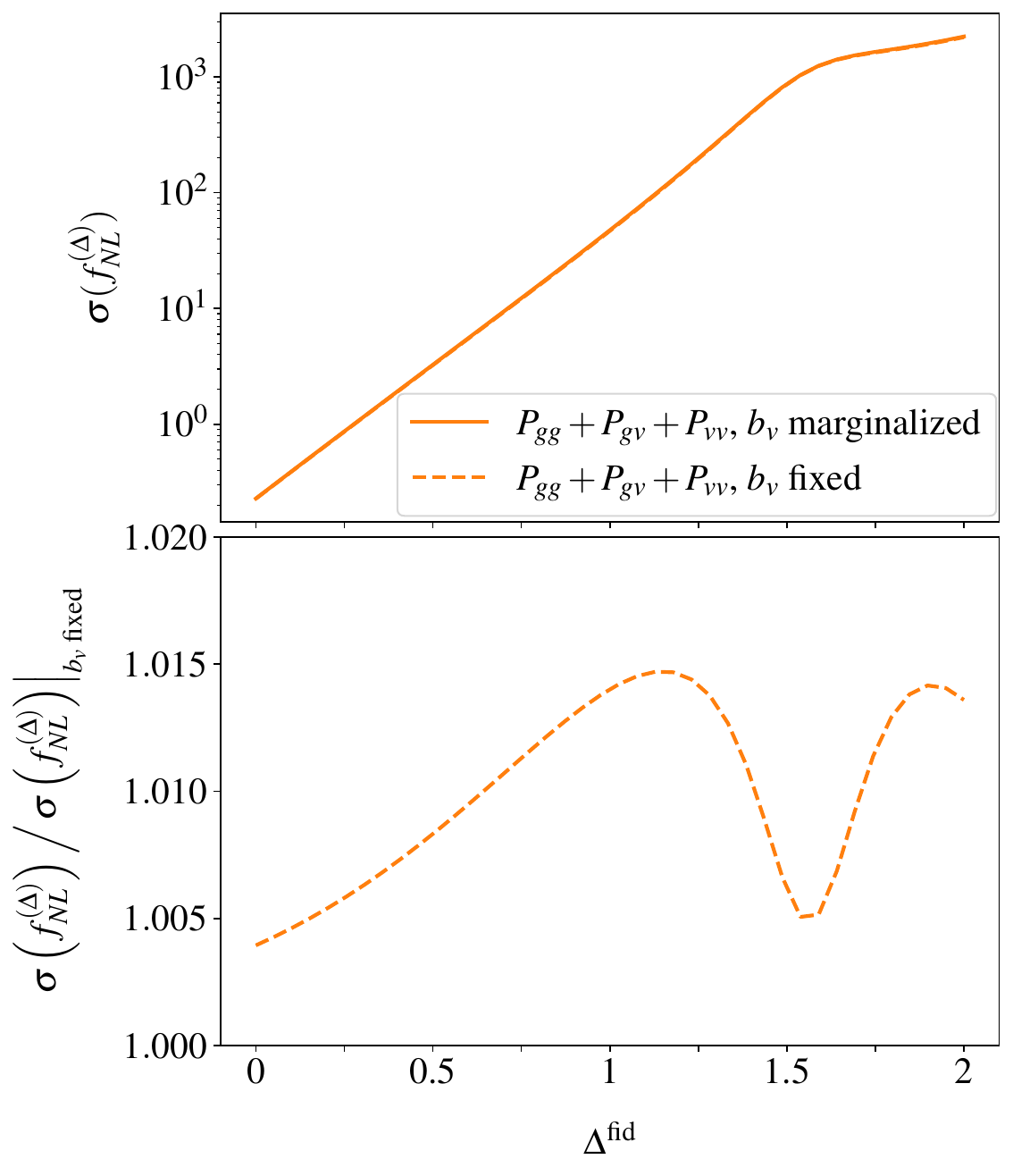}
    \caption{\textbf{Impact of optical depth degeneracy on $\sigma(\fnldelta)$.} Upper panel: we show how $1\sigma$ constraints on $\fnldelta$ are improved when the optical depth degeneracy is eliminated by fixing $b_v=1$ (dashed) compared to marginalizing over $b_v$ (solid), as a function of the fiducial value of the scaling exponent. Note these two curves lie nearly on top of one another. Lower panel: the ratio of $\sigma(\fnldelta)$ when $b_v$ is marginalized to when $b_v$ is fixed.}
    \label{fig:nobv_fixDelta}
\end{figure}

In Fig.~\ref{fig:nobv_fixDelta}, we demonstrate the impact that the optical depth degeneracy, e.g. uncertainty in the velocity reconstruction bias parameter $b_v$, has on the results we have presented here. In the top panel, we compare the $1\sigma$  $\fnldelta$ constraints from the combination of galaxy and kSZ tomography when marginalizing over $b_v$ to assuming a fixed value $b_v =1$. The bottom panel shows the fractional difference between these two curves, indicating that taking $b_v$ to be perfectly known leads to constraints on $\fnldelta$ that are overoptimistic by less than $\approx \! 2\%$. Hence, the optical depth degeneracy does not significantly diminish the ability of kSZ tomography to improve constraints on primordial non-Gaussianity via measurement of scale-dependent bias.

\subsubsection{Linear galaxy bias \label{subsubsec:b1}}
\begin{figure}
    \centering
\includegraphics[width=1.\columnwidth]{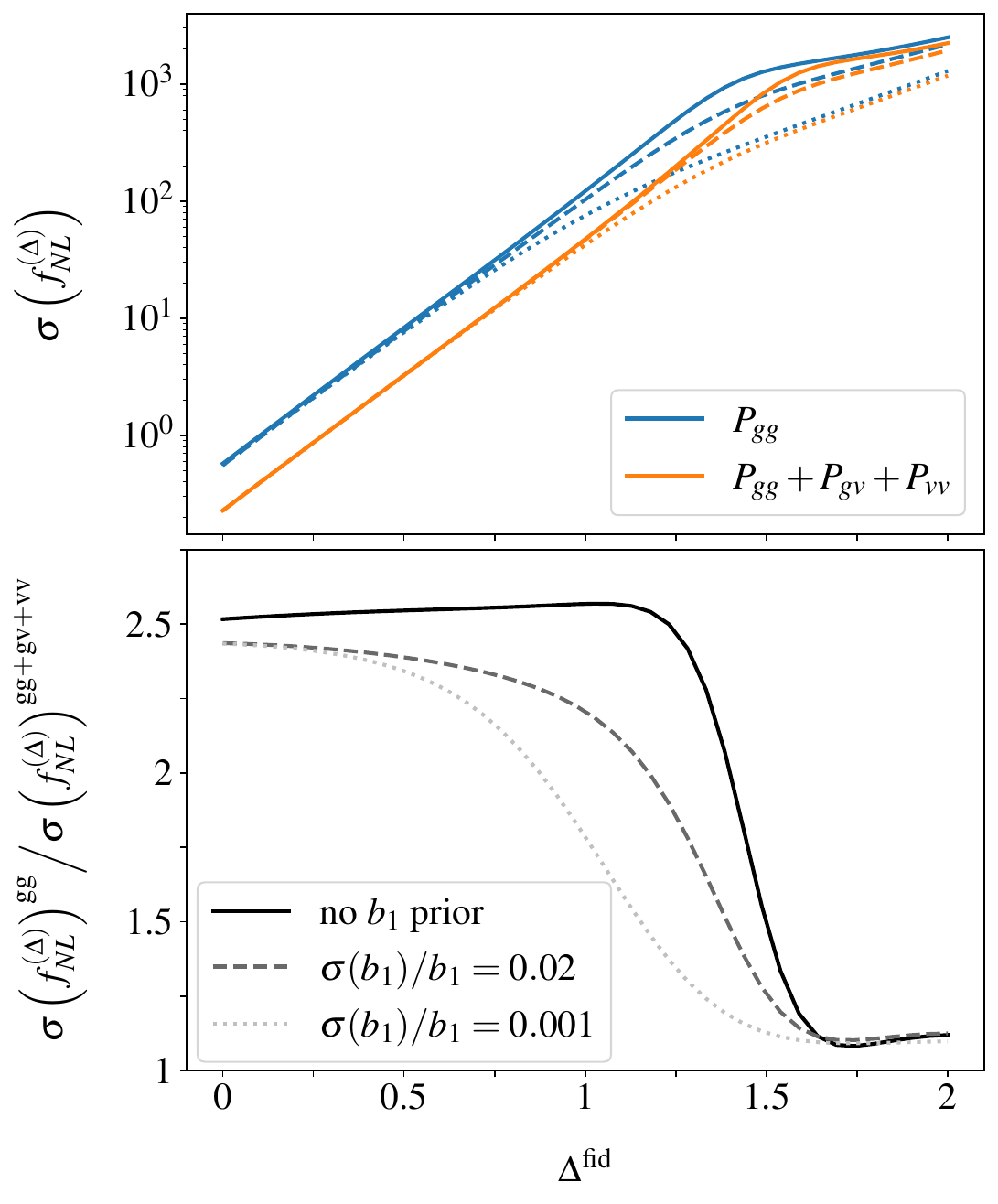}
    \caption{\textbf{Impact of various $b_1$ priors on $\fnldelta$ constraints.} Upper panel: $1\sigma$ constraints on $\fnldelta$ from galaxy (blue) and galaxy and kSZ (orange) tomography for no $b_1$ prior (solid), a $2\%$ prior (dashed), and a $0.1\%$ prior (dotted). Lower panel: ratio of $\sigma(\fnldelta)$ from galaxy tomography to $\sigma(\fnldelta)$ from galaxy and kSZ tomography for the different $b_1$ priors.}
    \label{fig:b1priors_fixDelta}
\end{figure}

In Fig.~\ref{fig:b1priors_fixDelta} we  demonstrate the impact of uncertainty in the linear bias by considering the effect that various $b_1$ priors have on constraints on $\fnldelta$. For simplicity we have chosen to implement the same priors across the five redshift bins. For illustrative purposes, we show a $2\%$ and a $0.1\%$ prior; the $2\%$ prior is roughly the precision needed to see significant degeneracy breaking in the feature at $\deltafid \approx 1.4$, while the $0.1\%$ is a rough `upper limit' beyond which a stronger prior does not impact the results. In next-generation surveys, achieving external, redshift-independent $b_1$ priors of this magnitude is in all likelihood unrealistic (although potentially some progress can be made, for example, by  cross correlating lensing with the same galaxy sample). However, these simplifying choices allow us to easily demonstrate the  effect that uncertainty in the linear galaxy bias has on $\sigma(\fnldelta)$ over the range of scaling exponents.

We find that uncertainty in $b_1$ significantly hinders constraints on $\fnldelta$ for scaling exponents $\deltafid \approx 1.5$, with the most significant effect occurring for slightly larger scaling exponents when including kSZ tomography.  Near this value of $\deltafid$, the two priors on $b_1$ have the potential to reduce $\sigma(\fnldelta)$ from galaxies alone by $50\%$ and over a factor of $3$ respectively. KSZ tomography helps to reduce the sensitivity to uncertainty in $b_1$, in part breaking the degeneracy between the linear bias and the non-Gaussianity amplitude, and hence the priors yield a smaller reduction in $\sigma(\fnldelta)$ from the combination of galaxiy and kSZ tomography. Correspondingly, the improvement in $\sigma(\fnldelta)$ from including kSZ tomography is generally less as the $b_1$ prior gets stronger. 



Even so, setting a strong enough prior such that $b_1$ is effectively fixed does not push down the measurement error enough to make galaxy and kSZ tomography competitive with the CMB at large scaling exponents. As we pointed out in the previous section, the practical application of kSZ tomography is likely to be for smaller scaling exponents,  $\deltafid \lesssim 0.7 - 0.85$, and here the impact of uncertainty in $b_1$ is very small, as evidenced by minimal response to the strong priors for those scaling exponents. In this range, for galaxies alone, the stronger prior reduces the measurement error on $\fnldelta$ by at most $\approx 30\%$, while with the addition of kSZ tomography, this figure drops to roughly $5\%$. Overall, these results suggest that the practical application of kSZ tomography to probing non-Gaussianity beyond the local type is robust to uncertainty in standard linear galaxy bias.

\subsubsection{Gradient biases \label{subsubsec:gradientbias}}
\begin{figure}
    \centering
\includegraphics[width=1.\columnwidth]{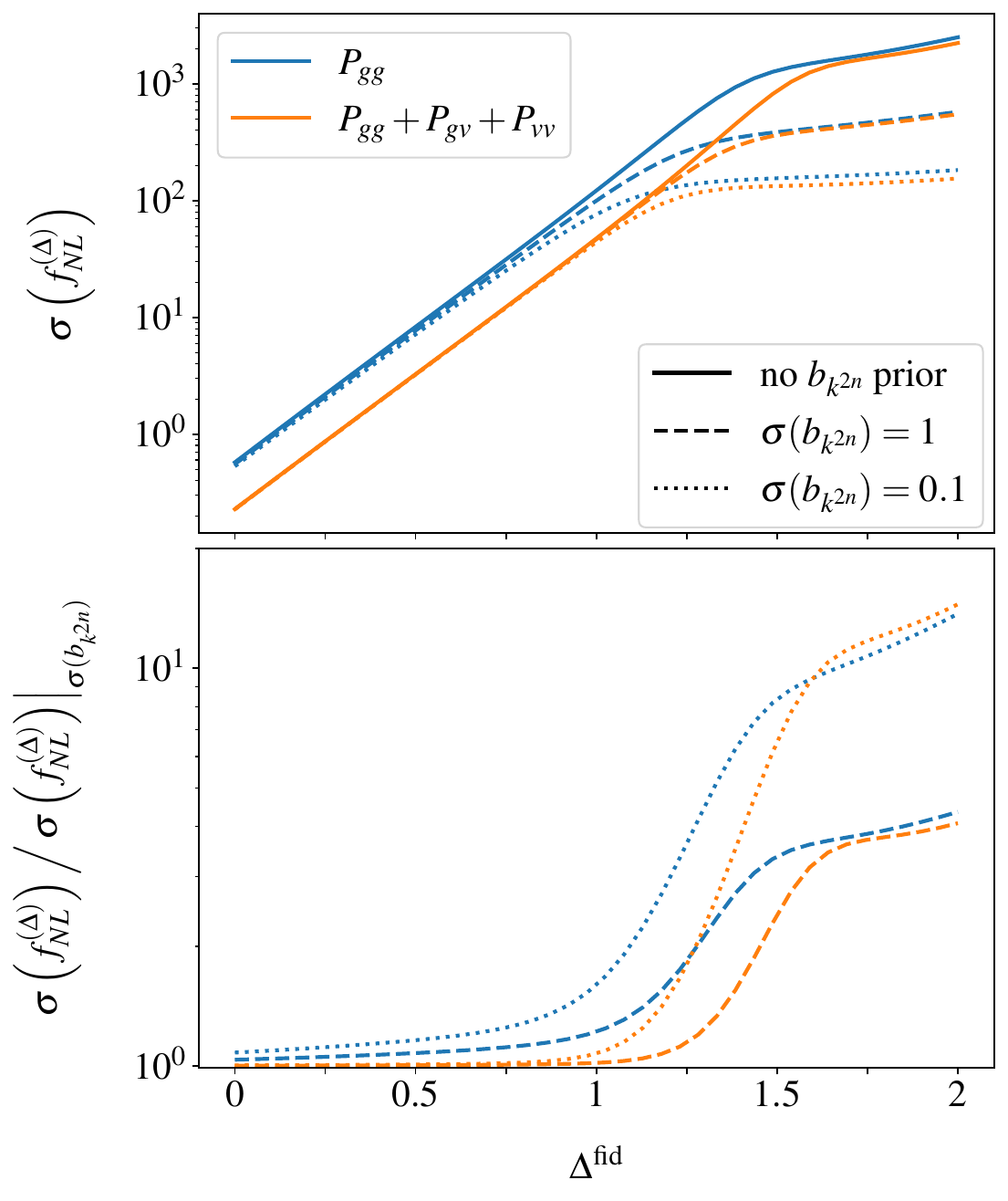}
    \caption{\textbf{Impact of gradient bias on $\fnldelta$ constraints.}  Upper panel: comparing 1$\sigma$ constraints on the non-Gaussianity amplitude from galaxy (blue) and galaxy and kSZ (orange) tomography for no $b_{k^{2n}}$ prior (solid), $\sigma(b_{k^{2n}}) =1 $ (dashed), and $\sigma(b_{k^{2n}}) = 0.1$ (dotted), over a range of fiducial scaling exponents. Lower panel: ratio of $\sigma(\fnldelta)$ with no $b_{k^{2n}}$ prior to $\sigma(\fnl)$ with $b_{k^{2n}}$ priors.  }
    \label{fig:bk2npriors_fixDelta}
\end{figure}

Our baseline forecasts have marginalized over the gradient biases $b_{k^2},b_{k^4}$ which modify the galaxy bias in a scale dependent fashion, roughly on a characteristic scale $k_* = 1/R_* \approx 0.25 \mathrm{Mpc}^{-1} $ (depending on the precise value of $b_{k^{2n}}$). On one hand, this is a much smaller scale than the regime where we expect kSZ tomography to add appreciable constraining power. On the other hand, for larger values of the scaling exponent, the impact of $\bng(k)$ shifts from larger to smaller scales and becomes more blue-tilted, and so heuristically we may expect the degeneracy between the gradient biases and $\bng(k)$ to be greater for larger $\deltafid$. We quantify the impact that uncertainty in the gradient biases has on constraints on $\fnldelta$ in Fig.~\ref{fig:bk2npriors_fixDelta}, where we show forecasted $1\sigma$ constraints on $\fnldelta$ with and without $b_{k^{2n}}$ priors. As we did with the linear bias, we take the simplifying approach of imposing the same prior for all redshift bins.\footnote{Achieving observational priors on small scale galaxy bias is very likely impractical in the near future. However, it is possible that theory priors, e.g. from analytic models or simulations, may provide some guidance \cite{Schmittfull:2014tca,Lazeyras:2015lgp,Desjacques:2016bnm}, and implicit in the gradient expansion is the assumption the $b_{k^{2n}}$ are order one parameters \cite{Gleyzes:2016tdh}.} As expected, priors on the the small scale gradient biases make a more significant impact on $\sigma(\fnldelta)$ at larger scaling exponents, for which the effect of $\bng(k)$ is more heavily weighted toward smaller scales. For scaling exponents near $\deltafid =2$, constraints on $\fnldelta$ from both galaxy and kSZ tomography improve significantly (by over an order of magnitude) for strong priors on the gradient biases, confirming significant degeneracy in that regime. As the lower panel of Fig.~\ref{fig:bk2npriors_fixDelta} indicates, the inclusion of kSZ tomography reduces the sensitivity to uncertainty in the gradient bias over most of the range  $\deltafid \in [0,2]$, although this behavior is somewhat nontrivial. In short, the inclusion of kSZ tomography helps to mildly break the degeneracy between the gradient bias and the non-Gaussianity amplitude.  

From a practical point of view, the impact of uncertainty in the gradient bias is very similar to that of the linear bias. Even effectively fixing the gradient biases does not reduce the measurement error enough at large scaling exponents to bring galaxy and kSZ tomography into competition with the CMB, while at scaling exponents ($\deltafid \lesssim 0.85$) for which galaxy and kSZ tomography are likely to outperform the CMB, uncertainty in $b_{k^{2n}}$ is not very significant. Similar to the case with the linear bias, the addition of kSZ tomography helps to effectively eliminate what moderate sensitivity the galaxy surveys have to uncertainty in $b_{k^{2n}}$: fixing $b_{k^{2n}}$ would reduce the measurement error by up to $30\%$ for galaxies alone, but the inclusion of kSZ tomography reduces this figure to under $3\%$. In summary, these results suggests that the near term, practical application of kSZ tomography for improving constraints on primordial non-Gaussianity will be largely robust to  uncertainty in the gradient bias, or likely any biases that only become important on non-linear scales.




\subsubsection{Cosmological parameters \label{subsubsec:effect_of_cosmology}}

\begin{figure}
\includegraphics[width=1.\columnwidth]{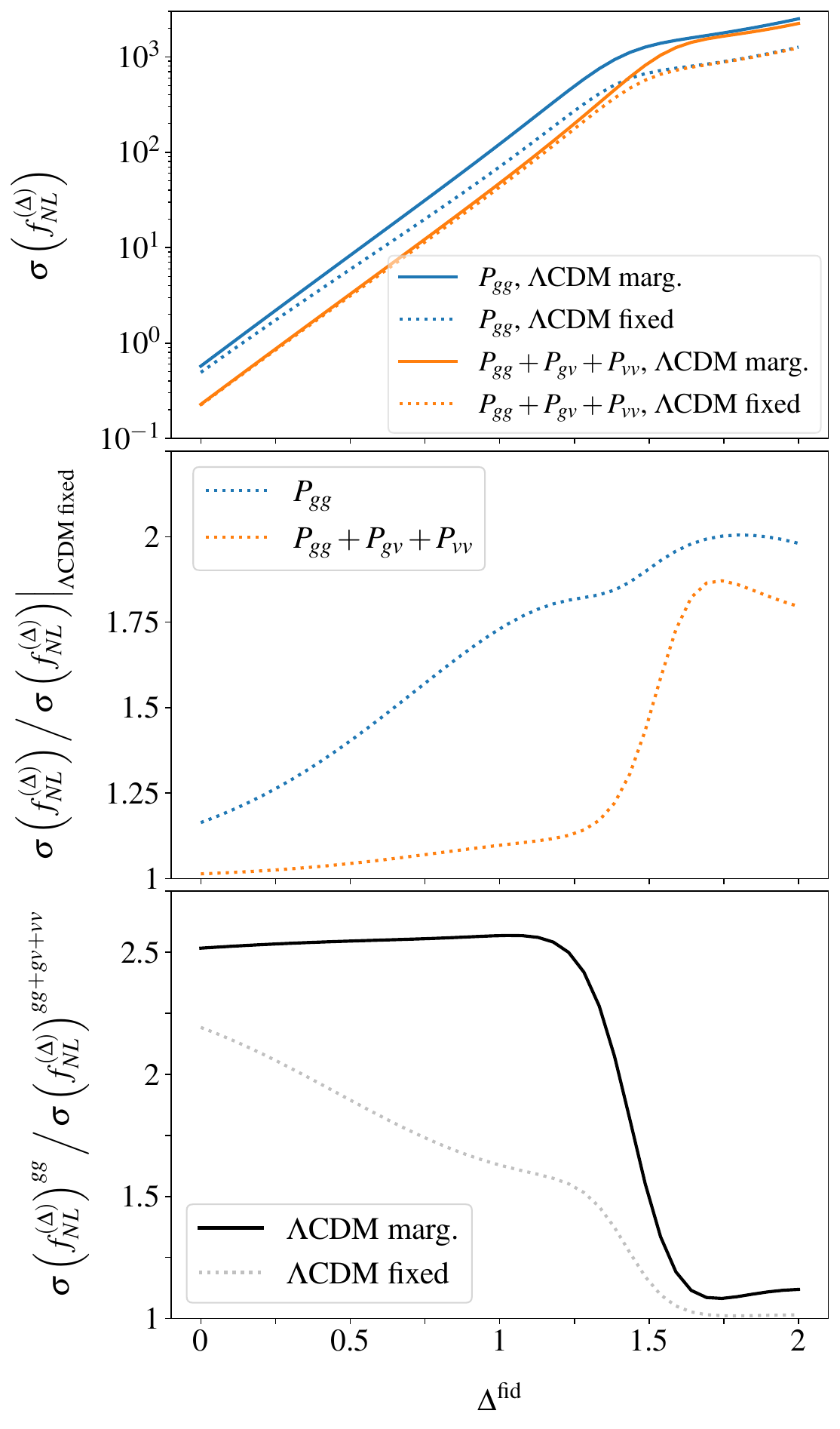}
    \caption{\textbf{Impact of uncertainty in $\Lambda$CDM parameters on $\sigma(\fnldelta)$.} Upper panel: we compare the forecasted $1\sigma$ constraints on the non-Gaussianity amplitude $\fnldelta$ from galaxy (blue) and galaxy and kSZ (orange) tomography when the $\Lambda$CDM parameters are marginalized (solid) versus when they are fixed (dotted). Middle panel: the ratio of $\sigma(\fnldelta)$ when the $\Lambda$CDM parameters are marginalized to $\sigma(\fnldelta)$ when they are fixed. Lower panel: the ratio of $\sigma(\fnldelta)$ from galaxy tomography to $\sigma(\fnldelta)$ from galaxy and kSZ tomography when the  $\Lambda$CDM parameters are marginalized (solidm black), or fixed (dotted, gray). }
    \label{fig:fixDelta_no_LCDM}
\end{figure}

In our baseline forecasts we have marginalized over a six-parameter $\Lambda$CDM model with fiducial values given in Eq.~(\ref{eq:LCDMfid}). Previous work has shown that, for the scaling exponent held fixed to $\Delta=0$, fixing the cosmology does not lead to significantly overoptimistic results because the degeneracy between $\fnlloc$ and the six $\Lambda$CDM parameters is sufficiently weak in the matter power spectrum \cite{Munchmeyer:2018eey}. In Fig.~\ref{fig:fixDelta_no_LCDM}  we demonstrate how a fixed cosmology affects constraints on $\fnldelta$. In agreement with previous work, we find that for fixed $\Delta=0$, fixing the cosmology only underestimates $\sigma(\fnldelta)$  from galaxies alone by about $15\%$, and the inclusion of kSZ tomography negates this effect. Hence in this case, the improvement in $\sigma(\fnlloc)$ gained from including kSZ tomography does not change very significantly when marginalizing over the six $\Lambda$CDM parameters. 

However, this conclusion changes if the scaling exponent deviates from $\Delta=0$. Fixing the cosmology can underestimate $\sigma(\fnldelta)$ by up to a factor of 2, depending on the fiducial value of the scaling exponent, mostly due to degeneracy between $\fnldelta$ and $n_s$, $\Omega_b h^2$, and $\Omega_c h^2$. The reduction in $\sigma(\fnldelta)$ from fixing the $\Lambda$CDM parameters is much greater for galaxies alone. That is, kSZ tomography significantly reduces the impact that uncertainty in the $\Lambda$CDM parameters has on constraints on $\fnldelta$. For example, for $\deltafid \lesssim 0.85$, the measurement error from galaxies alone is up to $60\%$ larger with $\pi^{\Lambda\mathrm{CDM}}$ marginalized, but less than $10\%$ larger with the inclusion of kSZ tomography.

For simplicity, and to compare with previous work, we have made the comparison with a fixed cosmology, but in practice the cosmological parameters will receive strong priors from, for example, the contemporaneous CMB experiment used in the kSZ velocity reconstruction. However, we find only $\mathcal{O}(10\%)$ difference in our results between fixing the $\Lambda$CDM parameters and using Planck priors \cite{Planck:2018vyg}, with no qualitative difference in the conclusions. Hence in Fig.~\ref{fig:fixDelta_no_LCDM}, the gray dotted curve in the bottom panel is likely a more realistic representation of the improvement in $\sigma(\fnldelta)$ one can expect from kSZ tomography. Likewise, in the upper panel of Fig.~\ref{fig:fixDelta_no_LCDM}, the dotted lines are more realistic predictions for the total 1$\sigma$ constraints on $\fnldelta$. This highlights that precise measurement of the underlying cosmology can be important for extracting constraints on primordial non-Gaussianity with scale-dependent galaxy bias. The upshot is that for the scaling exponents of practical interest, kSZ tomography will enable a roughly $2$-fold reduction in $\sigma(\fnldelta)$ over galaxies alone.

\subsubsection{Survey specifications \label{subsubsec:surveys}}
Our forecasts with galaxy and kSZ tomography have assumed an S4-like CMB experiment and an LSST-like galaxy survey. Here we consider how our primary conclusions are affected by some of the specifications of the experimental setup.

\begin{figure}
    \centering
\includegraphics[width=1.\columnwidth]{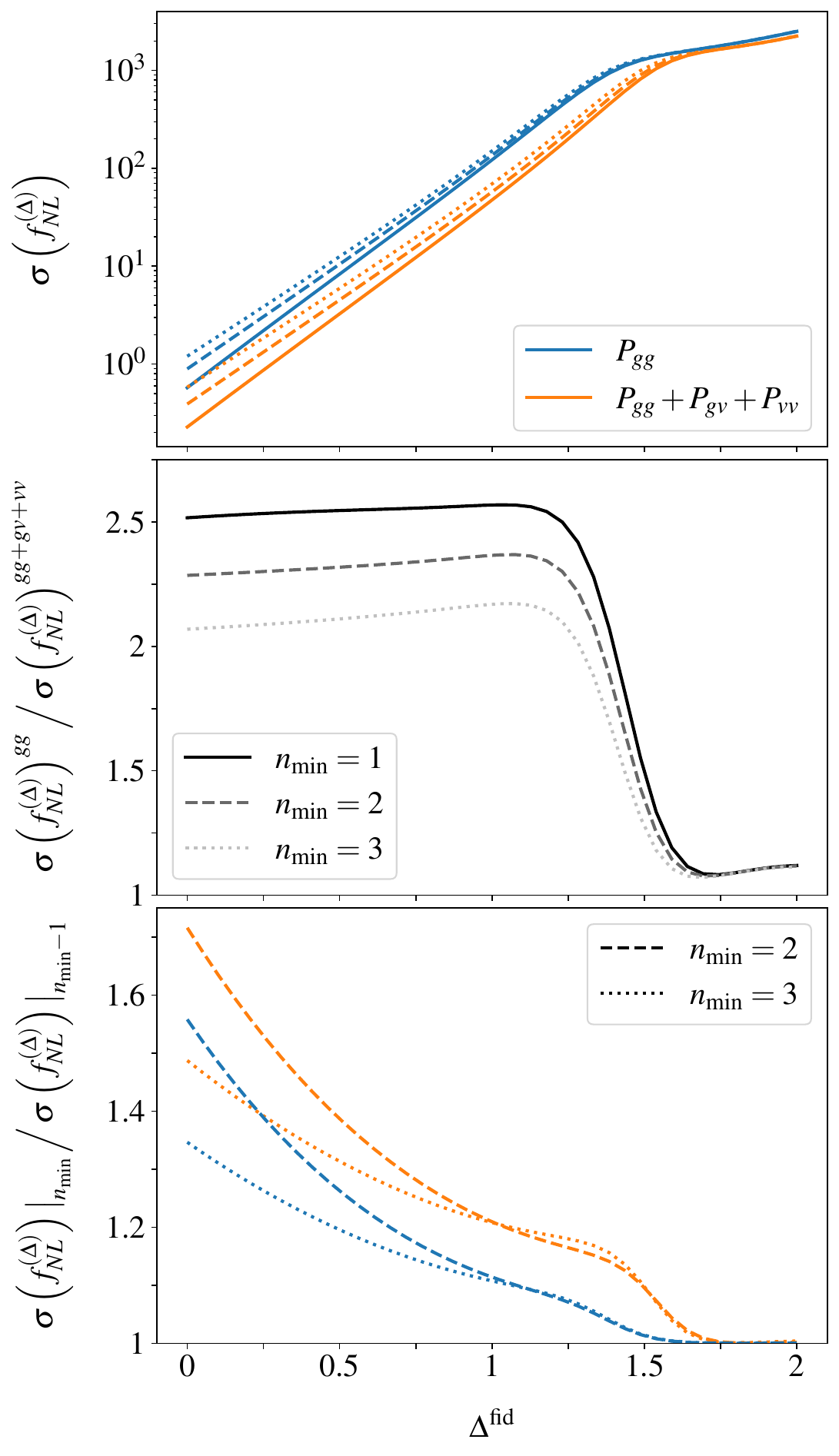}
    \caption{\textbf{Impact of $k_\mathrm{min}$ on $\fnldelta$ constraints from galaxy and kSZ tomography} Upper panel: $1\sigma$ constraints on $\fnldelta$  for a range of $k_{\mathrm{min}}$ values determined by the largest mode in the box, for galaxy tomography alone (blue) and galaxy and kSZ tomography (orange). Middle panel: ratio of $\sigma(\fnldelta)$ from galaxy tomography to $\sigma(\fnldelta)$ from galaxy and velocity tomography. Lower panel: ratio of $\sigma(\fnldelta)$ with different $k_{\rm{min}}$. In all panels, solid lines include all the modes in the box, dashed lines take $n_{\mathrm{min}}=2$, and dotted lines take $n_{\mathrm{min}}=3$. }
    \label{fig:kmin_fixDelta}
\end{figure}

First, we consider the consequences of a galaxy survey with access to fewer large scale modes. This is important for two reasons: first, for the linear galaxy power spectrum, the largest wavelength modes are measured with the less accuracy than modes close to the baryon acoustic oscillation scale or the peak of the matter power spectrum; secondly, velocity tomography is projected to be a powerful probe on the largest scales where it benefits from the $k^2$ signal-to-noise scaling discussed above in Sec.~\ref{sec:tomography}. In Fig.~\ref{fig:kmin_fixDelta}, we show the impact of varying the smallest scale accessible to the survey, $k_{\mathrm{min}}$, on $\sigma(\fnldelta)$, For simplicity, we vary $k_{\mathrm{min}}$ in the box formalism we work in by fixing $n_{\mathrm{min}}>1$, where the $n$th mode is $k_n = n\pi/(V^{1/3})$, and the baseline analysis sets $k_{\mathrm{min}}$ via $n_{\mathrm{min}}=1$. More realistically, these largest scale modes will be present but with significant noise, so our analysis here provides a simple upper bound on increase in measurement error. As the bottom panel of Fig.~\ref{fig:kmin_fixDelta} indicates, the overall constraints on $\fnldelta$ suffer significantly from the absence of the largest scale modes. Moving from $n_{\rm{min}}=1$ to $n_{\rm{min}}=2$ can increase the measurement error $\sigma(\fnldelta)$ by up to  $\approx 55\%$ ($70\%$) for galaxy (kSZ and galaxy) tomography, and moving from $n_{\rm{min}}=2$ to $n_{\rm{min}}=3$ causes a moderately smaller, but still significant, further increase. 
KSZ tomography suffers proportionally more from the loss of the largest modes, which is also reflected in how
the kSZ-driven improvement in the constraints suffers from the loss in large scale information, as can be seen  in the middle panel of Fig.~\ref{fig:kmin_fixDelta}. For scaling exponents $\deltafid \lesssim 1.4$, the improvement in $\sigma(\fnldelta)$ from kSZ is reduced by roughly $15\%$ for each removed mode. The impact is much less severe for larger values of the scaling exponent because the effect of $\bng(k)$ shifts to smaller scales, and so the loss of large scale information is less detrimental.

The galaxy number density $n_g$ is also crucial for obtaining strong constraints on $\fnldelta$ from scale dependent galaxy bias. For the case of $\fnlloc$, the improvement in $\sigma(\fnl)$ from kSZ tomography can be significantly better for larger galaxy number densities \cite{Munchmeyer:2018eey}. We do not carry out a similar test here, as the result for cases with scaling exponents close to $\Delta = 0$ should be very similar to the result for $\fnlloc$. For  significantly larger scaling exponents, where the effects of $\bng(k)$ move to smaller scales, a larger $n_{g}$ should not significantly improve constraints from kSZ tomography for two reasons. First, for an S4-like CMB experiment, the velocity reconstruction noise is approaching saturation with respect to $n_g$ \cite{Munchmeyer:2018eey}. Second, on smaller scales (where there is more information about larger scaling exponents), the velocity reconstruction noise is totally dominated by the exponential photo-$z$ error, rendering even significant increases in $n_g$ irrelevant. Indeed, as Fig.~\ref{fig:kmin_fixDelta} already demonstrates, changes in signal-to-noise at large scales has a diminishing effect as we go to larger scaling exponents.

\begin{figure}
    \centering
\includegraphics[width=1.\columnwidth]{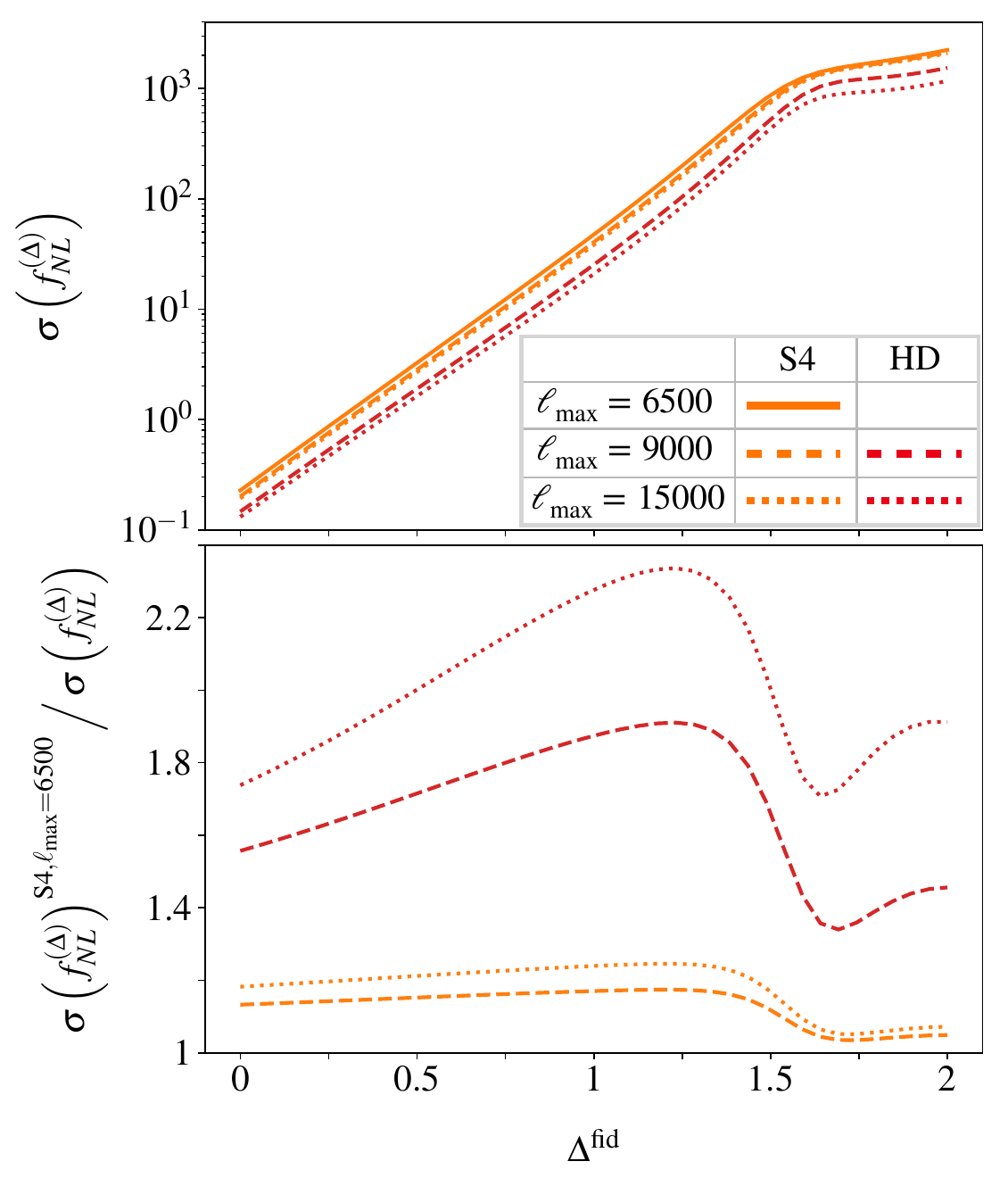}
    \caption{\textbf{Impact of a higher precision CMB experiment on $\fnldelta$ constraints from kSZ tomography.} Upper panel: forecasted $1\sigma$ constraints on $\fnldelta$ from the combination of galaxy and kSZ tomography assuming CMB S4 specifications (orange) and CMB HD specifications (red), with $\ell_{\mathrm{max}} = (6500,9000,15000)$ in (solid, dashed, dotted) lines. All curves assume an LSST-like galaxy survey as described in Sec.~\ref{sec:forecasts}. Lower panel: ratio of forecasted $1\sigma$ $\fnldelta$ constraints from the various CMB setups to our baseline S4 setup with $\ell_{\mathrm{max}} =6500$.}
    \label{fig:S4vsHD}
\end{figure}

KSZ tomography is also sensitive to the details of the CMB experiment, and in particular the velocity reconstruction benefits from a high precision CMB experiment that can accurately measure the temperature anisotropies at high multipole $\ell$. In Fig.~\ref{fig:S4vsHD} we examine the impact of increasing the maximum multipole in the CMB survey for our baseline S4+LSST setup. We also compare these results to a futuristic scenario with a CMB experiment with specifications matching that of CMB HD \cite{Sehgal:2019ewc}, specifically with a $\sim 20^{\prime \prime}$ beam and white noise $\Delta_T = 0.1 \, \mu \mathrm{K}^{\prime}$. We find that the reduction in $\sigma(\fnldelta)$ from and S4-like survey with a more optimistic maximum multipole, $\ell_{\mathrm{max}}=9000$, yields a nearly $15\%$ smaller $\sigma(\fnldelta)$ for $\deltafid \lesssim 1.3$. A futuristic, HD-like CMB experiment would offer significantly stronger constraints. For example, with $\ell_{\mathrm{max}} =15000$, the improvement over our baseline S4 results is $\gtrsim 75\%$. At large $\deltafid$, this improvement is not enough to bring kSZ and galaxy tomography into contention with the constraining power of the primary CMB itself, but constraints on $\fnlloc$, for example, would be much stronger, and such a setup would modestly increase the range of order unity constraints on $\fnldelta$ to $\deltafid \lesssim 0.4$. The additional constraining power comes from a reduction in the kSZ reconstruction noise that, on the largest scales, and at higher redshifts, is over an order of magnitude. Note that in this comparison, we have only changed the CMB specifications and not the details of the galaxy survey. This allows us to clearly isolate the impact that a higher precision CMB experiment has on constraining primordial non-Gaussianity with kSZ tomography. However, in practice, CMB HD would likely be accompanied by a contemporaneous galaxy survey, such as MegaMapper \cite{Schlegel:2019eqc}, and such a futuristic galaxy survey with improved number density and/or lower (or no) photo-$z$ error would benefit both galaxy and kSZ tomography, which may change the quantitative conclusions about how much improvement kSZ tomography offers in a futuristic setting. Our point here is simply to demonstrate that constraints on $\fnldelta$ from kSZ tomography can significantly benefit from a higher precision CMB experiment. 

\section{Discussion \label{sec:discussion}}

In this paper we have explored the possibility of using kSZ tomography to improve constraints on primordial non-Gaussianity beyond the local type. We have focused on a scenario where the light degrees of freedom present during inflation lead to a bispectrum scaling as $k^\Delta$ in the squeezed limit. This leaves a distinct imprint on the scale dependence of the galaxy bias, which can differ from the characteristic large scale $k^{-2}$ behavior associated with local-type primordial non-Gaussianity. 
The primary results of this paper, encapsulated in Figs.~\ref{fig:ksz_vs_gal_fnl_fixedDelta}-\ref{fig:ksz_vs_gal_Delta_varyDelta}, indicate that with next-generation CMB and galaxy surveys, kSZ tomography will significantly improve the measurement of the  scale-dependent galaxy bias on large scales for a range of scaling exponents $\Delta$, thereby tightening constraints on primordial non-Gaussianity beyond the local type. 

The main conclusion from this work is that the practical application of kSZ tomography for probing primordial non-Gaussianity is not limited to constraining $\fnlloc$. Most importantly, our results show that kSZ tomography will significantly improve constraints around $\fnldelta =0$ compared to galaxy surveys alone, enabling late-time probes of the matter power spectrum to outperform the CMB for a wider range of scaling exponents beyond $\Delta=0$. Similarly, kSZ tomography will extend the range of scaling exponent for which the thoeretical target of order unity constraints on $\fnl$ are achievable, opening a wider window into the physics of the early Universe. Furthermore, we have demonstrated that if a non-vanishing $\fnl$ were to be detected, kSZ tomography will improve the synergy between the CMB and late-time probes of primordial non-Gaussianity, enhancing the detection significance of $\fnldelta$ and significantly reducing the measurement error on the scaling exponent $\Delta$.  

In our analysis we have explored how improved constraints on primordial non-Gaussianity from kSZ tomography can quantitatively change in response to a range of potential biases, degeneracies, experimental parameters, and modelling details. However, we have demonstrated that our primary conclusions are robust to these various uncertainties. The optical depth degeneracy, which can potentially hinder cosmological inference with kSZ tomography, is essentially a non-factor in the context of the present work. This is a simple consequence of two factors. First, the key measurement is that of the large-scale galaxy bias, and second, combining galaxy and kSZ tomography is effectively a multi-tracer approach to measuring the total matter power spectrum. Combined, these mean that the main role of kSZ tomography in this context is to facilitate sample variance cancellation, circumventing cosmic variance limitations that would otherwise hinder the galaxy or velocity power spectrum in isolation. The combination of galaxy and kSZ tomography, therefore, enables the unambiguous measurement of the galaxy bias on the largest cosmological scales. In a similar vein, it was shown in Ref.\ \cite{Green:2023uyz} that, for an LSST-like survey, the constraints on $\fnldelta$ may significantly improve with explicitly multi-tracer galaxy tomography  (for beyond local-type primordial non-Gaussianity), while in Ref. \cite{Munchmeyer:2018eey} it was demonstrated that the additional constraining power for $\fnlloc$ offered by kSZ tomography depends on the multi-tracer nature of the galaxy survey. Hence, one important future direction is to explore the role kSZ tomography would play in constraining $\fnldelta$ if the galaxy surveys benefit from some internal sample variance cancellation.

We have also examined how uncertainty in the details of galaxy bias may impact constraints on primordial non-Gaussianity in this scenario. In particular, we have shown that uncertainty in both the standard linear galaxy bias $b_1$, as well as in the small scale biasing details (here studied via gradient biases $b_{k^{2n}}$) may significantly worsen constraints on $\fnldelta$, but that kSZ tomography generally helps to mitigate this effect. Furthermore, the degeneracies between the non-Gaussianity amplitude and the other contributions to galaxy bias are weakest for smaller scaling exponents, for which kSZ tomography holds the most promise. Therefore, in practice, we anticipate the uncertainty in the model for the galaxy bias will play little role where kSZ (and galaxy) tomography has any practical application. 

Beyond the details of biasing, our results show that the inclusion of kSZ tomography can significantly reduce the sensitivity of the measurement error on $\fnldelta$  to uncertainties in the $\Lambda$CDM parameters. This reduction is significant for non-Gaussianity beyond the local limit, particularly for scaling exponents $\Delta \approx 1$.  As a corollary, this means marginalizing over the cosmological model mildly overestimates the improvement in constraining power from kSZ tomography compared to galaxies in isolation. In real analysis, additional information about the cosmological model will be furnished by the CMB. In that case, a good rule of thumb is that, for scaling exponents for which the late-time power spectrum probes compete with the sensitivity of the CMB, kSZ tomography will reduce $\sigma(\fnldelta)$ from galaxies alone by about a factor of 2. These results indicate that the assistance granted by kSZ tomography in this context is not only to facilitate sample variance cancellation, but also to better pin down the cosmology that controls the underlying matter power spectrum, and that these cosmological parameters do introduce non-negligible degeneracies that diminish  constraints on the amplitude of primordial non-Gaussianity. This also suggests that future studies should examine, for example, how extended cosmological models, e.g. dynamical dark energy, impact constraints on primordial non-Gaussianity from measurements of the galaxy and velocity fields.

While we have focused on galaxy surveys and kSZ tomography, measurements of the CMB anisotropies place independent constraints on primordial non-Gaussianity. Although quasi-single-field shapes with scaling exponents $\Delta \gtrsim 0.85$, will very likely remain better probed by the CMB itself, our results reinforce findings that galaxy and kSZ tomography will play an important role in providing competitive, and possibly stronger, constraints on non-Gaussianity of the local type and beyond.  Fully characterizing how constraints from LSS will faithfully compare with future CMB analyses of quasi-single-field scenarios requires a more dedicated study, which we leave for future work.

While we have focused on probing primordial non-Gaussianity, the present work also highlights the broader capability of kSZ tomography as a test of any physics imprinted in scale-dependent bias on the largest cosmological scales. For example, simulations suggest that scale-dependent bias on large scales is present even in standard cosmological settings without primordial non-Gaussianity \cite{Springel:2017tpz}.  
Probing primordial non-Gaussianity, of local type and beyond, is one demonstration of the myriad new opportunities for cosmological inference that will be made possible by the greatly improved small scale sensitivity of next generation surveys. These surveys will open new avenues for probes of fundamental physics beyond the cosmic variance limit and robust to various astrophysical and cosmological uncertainties.

\section{Acknowledgements}
We are grateful to Gil Holder, Selim Hotinli, and Srinivasan Raghunathan for useful conversations. The authors are supported in part by the
United States Department of Energy, DE-SC0015655.

\bibliography{main}

\end{document}